\documentstyle[12pt,graphicx,pstricks,braket,amsmath,amssymb,enumitem,
                    epstopdf,pdfpages,cite,tabu,hyperref,subfigure]{article}
\numberwithin{equation}{section}

\newcommand{\hi}[1]{}

\newcommand{\FbR}{{\bar F}_R}
\newcommand{\Db}{{\bar D}}
\newcommand{\Sb}{{\bar S}}
\newcommand{\phib}{{\bar \phi}}
\newcommand{\FL}{{F}_L}
\newcommand{\FR}{{F}_R}
\newcommand{\oh}{\frac{1}{2}}
\begin{document}

\def\AEF{A.E. Faraggi}

\def\JHEP#1#2#3{{\it JHEP} {\textbf #1}, (#2) #3}
\def\vol#1#2#3{{\bf {#1}} ({#2}) {#3}}
\def\NPB#1#2#3{{\it Nucl.\ Phys.}\/ {\bf B#1} (#2) #3}
\def\PLB#1#2#3{{\it Phys.\ Lett.}\/ {\bf B#1} (#2) #3}
\def\PRD#1#2#3{{\it Phys.\ Rev.}\/ {\bf D#1} (#2) #3}
\def\PRL#1#2#3{{\it Phys.\ Rev.\ Lett.}\/ {\bf #1} (#2) #3}
\def\PRT#1#2#3{{\it Phys.\ Rep.}\/ {\bf#1} (#2) #3}
\def\MODA#1#2#3{{\it Mod.\ Phys.\ Lett.}\/ {\bf A#1} (#2) #3}
\def\RMP#1#2#3{{\it Rev.\ Mod.\ Phys.}\/ {\bf #1} (#2) #3}
\def\IJMP#1#2#3{{\it Int.\ J.\ Mod.\ Phys.}\/ {\bf A#1} (#2) #3}
\def\nuvc#1#2#3{{\it Nuovo Cimento}\/ {\bf #1A} (#2) #3}
\def\RPP#1#2#3{{\it Rept.\ Prog.\ Phys.}\/ {\bf #1} (#2) #3}
\def\APJ#1#2#3{{\it Astrophys.\ J.}\/ {\bf #1} (#2) #3}
\def\APP#1#2#3{{\it Astropart.\ Phys.}\/ {\bf #1} (#2) #3}
\def\EJP#1#2#3{{\it Eur.\ Phys.\ Jour.}\/ {\bf C#1} (#2) #3}
\def\etal{{\it et al\/}}
\def\notE6{{$SO(10)\times U(1)_{\zeta}\not\subset E_6$}}
\def\E6{{$SO(10)\times U(1)_{\zeta}\subset E_6$}}
\def\highgg{{$SU(3)_C\times SU(2)_L \times SU(2)_R \times U(1)_C \times U(1)_{\zeta}$}}
\def\highSO10{{$SU(3)_C\times SU(2)_L \times SU(2)_R \times U(1)_C$}}
\def\lowgg{{$SU(3)_C\times SU(2)_L \times U(1)_Y \times U(1)_{Z^\prime}$}}
\def\SMgg{{$SU(3)_C\times SU(2)_L \times U(1)_Y$}}
\def\Uzprime{{$U(1)_{Z^\prime}$}}
\def\Uzeta{{$U(1)_{\zeta}$}}

\newcommand{\cc}[2]{c{#1\atopwithdelims[]#2}}
\newcommand{\bev}{\begin{verbatim}}
\newcommand{\beq}{\begin{equation}}
\newcommand{\ba}{\begin{eqnarray}}
\newcommand{\ea}{\end{eqnarray}}

\newcommand{\beqa}{\begin{eqnarray}}
\newcommand{\beqn}{\begin{eqnarray}}
\newcommand{\eeqn}{\end{eqnarray}}
\newcommand{\eeqa}{\end{eqnarray}}
\newcommand{\eeq}{\end{equation}}
\newcommand{\beqt}{\begin{equation*}}
\newcommand{\eeqt}{\end{equation*}}
\newcommand{\Eev}{\end{verbatim}}
\newcommand{\bec}{\begin{center}}
\newcommand{\eec}{\end{center}}
\newcommand{\bes}{\begin{split}}
\newcommand{\ees}{\end{split}}
\def\ie{{\it i.e.~}}
\def\eg{{\it e.g.~}}
\def\half{{\textstyle{1\over 2}}}
\def\nicefrac#1#2{\hbox{${#1\over #2}$}}
\def\third{{\textstyle {1\over3}}}
\def\quarter{{\textstyle {1\over4}}}
\def\m{{\tt -}}
\def\mass{M_{l^+ l^-}}
\def\p{{\tt +}}

\def\slash#1{#1\hskip-6pt/\hskip6pt}
\def\slk{\slash{k}}
\def\GeV{\,{\rm GeV}}
\def\TeV{\,{\rm TeV}}
\def\y{\,{\rm y}}

\def\l{\langle}
\def\r{\rangle}
\def\LRS{LRS  }

\begin{titlepage}
\samepage{
\setcounter{page}{1}
\rightline{LTH--1085}
\vspace{1.5cm}

\begin{center}
 {\Large \bf 
LHC di--photon excess \\ \medskip 
and Gauge Coupling Unification in \\ \medskip 
Extra $Z^\prime$ Heterotic--String Derived Models}
\end{center}

\begin{center}

{\large
J. Ashfaque$^\clubsuit$\footnote{email address: jauhar@liverpool.ac.uk}, 
L. Delle Rose$^\spadesuit$\footnote{email address: l.delle-rose@soton.ac.uk},  
A.E. Faraggi$^\clubsuit$\footnote{
		                  email address: alon.faraggi@liv.ac.uk}
 and
C. Marzo$^\diamondsuit$\footnote{email address: carlo.marzo@le.infn.it}
}\\
\vspace{1cm}
$^\clubsuit${\it  Dept.\ of Mathematical Sciences,
             University of Liverpool,
         Liverpool L69 7ZL, UK\\}
\vspace{.025in}
$^\spadesuit${\it School of Physics and Astronomy, 
             University of Southampton, \\
         Southampton SO17 1BJ, UK\\
         Dept.\ of Particle Physics, 
         Rutherford Appleton Laboratory, \\
         Chilton, Didcot, OX11 0QX, UK \\
         }
\vspace{.025in}
$^\diamondsuit${\it  Dept.\ di Matematica e Fisica ``Ennio De Giorgi'',
             Universit\`a del Salento and INFN-Lecce,
         Via Arnesano, 73100 Lecce, IT\\}

\end{center}

\begin{abstract}
A di--photon excess at the LHC can be explained 
as a Standard Model singlet that is produced and decays by 
heavy vector--like colour triplets and electroweak doublets
in one--loop diagrams. The characteristics of the required 
spectrum are well motivated in heterotic--string constructions
that allow for a light $Z^\prime$. Anomaly cancellation of the
$U(1)_{Z^\prime}$ symmetry requires the existence of the Standard 
Model singlet and vector--like states in the vicinity of the 
$U(1)_{Z^\prime}$ breaking scale. In this paper we show that 
the agreement with the gauge coupling data at one--loop
is identical to the case of the Minimal Supersymmetric Standard Model, 
owing to cancellations between the additional states. We further
show that effects arising from heavy thresholds may push 
the supersymmetric spectrum beyond the reach of the LHC, 
while maintaining the agreement with the gauge coupling data. 
We show
that the string inspired model can indeed produce an
observable signal and discuss the feasibility of obtaining viable 
scalar mass spectrum.

\end{abstract}
\smallskip}
\end{titlepage}

\section{Introduction}

The Standard Model of particle physics provides viable parameterisation for 
all subatomic data to date. The most striking feature of the Standard Model, 
augmented by right--handed neutrinos that are required by the neutrino
data, is the embedding of its chiral spectrum in three chiral 16 
representations of $SO(10)$. Heterotic--string models 
give rise to spinorial 16 representations 
in the perturbative spectrum and therefore  
preserve the $SO(10)$ embedding of the 
Standard Model states \cite{heterotic, candelas}. 

Recently, a possible signal has been reported by the ATLAS \cite{atlas}
and CMS \cite{cms} 
collaborations that would indicate a clear deviation from the 
Standard Model. Both experiments reported early indications for 
enhancement of di--photon events with a resonance at 750GeV, 
and generated substantial interest \cite{flurry}. 
A plausible explanation for this enhancement 
is obtained if the resonant state is assumed to be a Standard Model
singlet state, and the production and decay are mediated by
heavy vector--like quark and lepton states
\cite{flurry, vectorlikeprops, alternatives}. These  
characteristics arise naturally in heterotic--string 
models that allow for a light extra $Z^\prime$ \cite{frdiphoton}. 

We note that the construction of heterotic--string models that
allow for a light $Z^\prime$ is highly non--trivial
\cite{zpbml, zprimeffm, frzprime} . The reason
being that the extra family universal $U(1)$ symmetries that are
typically discussed in the string--inspired literature tend 
to be anomalous and are therefore broken near the 
string scale \cite{au1}. The relevant symmetries 
tend to be anomalous due to the symmetry breaking
pattern $E_6\rightarrow SO(10)\times U(1)_\zeta$, induced at the 
string level by the Gliozzi--Scherk--Olive (GSO) projection \cite{gso}. 
In ref. \cite{frzprime} we used
the spinor--vector duality property of $Z_2\times Z_2$ 
orbifolds \cite{svd1,svd2} 
to construct a string derived model 
with anomaly free $U(1)_\zeta$, thus enabling it to remain
unbroken down to low scales. 

An additional constraint imposed by the heterotic--string is that the 
gauge, as well as the gravitational, 
couplings are unified at the string scale \cite{kaplu}.
Since the early nineties, much of the research 
on the phenomenology of supersymmetric
grand unified theories has been motivated 
by the observation that the unification of the gauge couplings in 
SUSY GUTs is compatible with the measured gauge coupling data
at the electroweak scale, provided that
we  assume that the spectrum between the 
two scales consists of that of the Minimal Supersymmetric 
Standard Model (MSSM) \cite{susyuni}. Following Witten we may assume that the 
string and GUT scales may coincide in the framework of $M$--theory
\cite{witten}. 

A vital question therefore is to examine what is the corresponding 
situation in the heterotic--string derived $Z^\prime$ models.
We find that, quite 
remarkably, in the $Z^\prime$ models 
the compatibility of gauge coupling unification
with the data at the electroweak scale
is identical to the case of the MSSM. 
We further
show that effects arising from heavy thresholds may push 
the supersymmetric spectrum beyond the reach of the LHC, 
while maintaining the agreement with the gauge coupling data. 
We show
that the string inspired model can indeed account for the 
observed signal and discuss the feasibility of obtaining viable 
scalar mass spectrum. 

While further data from the LHC did did not substantiate the 
observation of the di--photon excess \cite{atlasaug,cmsaug}, 
a di--photon excess is a general signature of this class 
of $Z^\prime$ models. The results presented in this paper 
are therofore relevant for continuing $Z^\prime$ searches
at the LHC. 

\section{The string model and extra $Z^\prime$} 

The difficulty in constructing heterotic--string models 
with light $Z^\prime$ symmetries arises due to the 
breaking of the observable $E_6$ symmetry in the string 
constructions by discrete Wilson lines to $SO(10)\times U(1)_\zeta$.
Application of the symmetry breaking at the string level 
results in the projection of some states from the physical 
spectrum. The consequence is that $U(1)_\zeta$ is in general
anomalous in the string vacua, and cannot remain unbroken to 
low scales. The extra $U(1)$ symmetry which is embedded in 
$SO(10)$, and is orthogonal to the Standard Model weak hypercharge,
is typically broken at the high scale to generate sufficiently
light neutrino masses. Flavour non--universal $U(1)$ symmetries 
must be broken above the deca--TeV scale to avoid conflict
with Flavour Changing Neutral Current (FCNC) constraints 
\cite{velduis}. 

The string derived model of ref. \cite{frzprime} was 
constructed in the free fermionic formulation \cite{fff} of the 
heterotic--string. The details of the construction, 
the massless spectrum of the model and its superpotential
are given in ref. \cite{frzprime} and will not be repeated here. 
We review here the properties of the model that are 
relevant for the anomaly free extra $Z^\prime$ symmetry.

The model utilises the spinor--vector duality symmetry
that was observed in the space of fermionic $Z_2\times Z_2$
orbifold compactifications \cite{svd1,svd2}. The spinor 
vector duality operates under exchange of the total number 
of spinorial $({16}\oplus\overline{16})$ representations of
$SO(10)$ with the total number of vectorial $10$ representations. 
For every string vacuum with a $\#_1$ of
$({16}\oplus\overline{16})$ representations and 
$\#_2$ of 10 representations there is a dual vacuum 
in which $\#_1\leftrightarrow\#_2$. The understanding of this
duality is facilitated by considering the vacua in which 
the $SO(10)\times U(1)_\zeta$ symmetry is enhanced to $E_6$. 
The chiral representations of $E_6$ are the $27$ and 
$\overline{27}$ and their decomposition under 
$SU(10)\times U(1)_\zeta$ is 
\beqn
27 & = & 16_{+{1/2}} + 10_{-1} + 1_{+2},\nonumber\\
\overline{27} & = & \overline{16}_{-{1/2}} + 10_{+1} + 1_{-2},\nonumber
\eeqn
where the subscript denotes the $U(1)_\zeta$ charge.
Thus, the string vacua with $E_6$ symmetry 
are self--dual with respect to the spinor--vector 
duality, {\it i.e.} in these vacua 
$\#_1(16\oplus\overline{16})=\#_2(10)$.
In this case $U(1)_\zeta$ is anomaly free 
by virtue of its embedding in $E_6$. 
There exist a discrete Wilson line that reduce
$E_6$ symmetry to $SO(10)\times U(1)_\zeta$ 
with $\#_1(16\oplus\overline{16})~ \& ~\#_2(10)$, 
and a corresponding discrete Wilson line with
 $\#_2(16\oplus\overline{16})~ \&~ \#_1(10)$ \cite{svd2}.

The string vacua with enhanced $E_6$ symmetry
correspond to heterotic--string vacua with 
$(2,2)$ worldsheet supersymmetry. We can realise
the $E_6$ symmetry by breaking the ten
dimensional untwisted gauge symmetry 
to $SO(8)^4$ \cite{svd1}. One of the $SO(8)$ factors 
is reduced further to $SO(2)^4$ and
the $E_6$ symmetry is generated from additional
sectors in the string vacua. In parallel to the spectral flow
operator on the supersymmetric side of the heterotic--string
that maps between different spacetime spin representations, 
there exists a spectral flow operator on the bosonic side. 
In the vacua with enhanced $E_6$ symmetry the spectral 
flow operator exchanges between the spinorial and vectorial
components in the $E_6$ representations. The spectral flow
operator is the $U(1)$ generator of the $N=2$ worldsheet 
supersymmetry on the bosonic side of the heterotic--string. 
In the vacua with broken $E_6$ symmetry, the $N=2$ worldsheet
supersymmetry on the bosonic side is broken and the spectral flow
operator induces the map between the spinor--vector dual vacua.
The picture was extended to other internal CFTs in ref. \cite{panos}.  

The class of $Z_2\times Z_2$ vacua affords another possibility.
It is possible to construct self--dual vacua with 
$\#_1(16\oplus\overline{16})=\#_2(10)$, without enhancing the 
gauge symmetry to $E_6$. This is the case if the different
components of the $E_6$ representations are obtained from 
different fixed points of the $Z_2\times Z_2$ orbifold. 
The spectrum then forms complete $E_6$ representations, 
but the gauge symmetry is not enhanced to $E_6$ and 
remains $SO(10)\times U(1)_\zeta$, with $U(1)_\zeta$ 
being anomaly free due to the fact that the chiral
spectrum still forms complete $E_6$ multiplets. It is 
important to note that this is possible only because
the spinorial and vectorial $SO(10)$ representations
are obtained from different fixed points.
Obtaining the $16$ and $(10+1)$ components at the same 
fixed point necessarily implies that the gauge 
symmetry is enhanced to $E_6$. 

The construction of ref. \cite{frzprime} utilises the
classification methods developed in ref. \cite{gkr} for
type IIB string and in ref. \cite{fknr}
for heterotic--string vacua with unbroken $SO(10)$
gauge group. The heterotic--string classification
was extended to vacua with the Pati--Salam 
and flipped $SU(5)$ subgroups of $O(10)$ in
refs. \cite{psclass} and \cite{fsu5class}, respectively.
In this method a space of the order of $10^{12}$ is spanned 
and models with specific phenomenological 
characteristics can be extracted. The string vacuum with
anomaly free $U(1)_{Z^\prime}$ is obtained by first 
trawling a self--dual $SO(10)$ model with 
six chiral families and subsequently breaking the 
$SO(10)$ symmetry to the Pati--Salam subgroup \cite{frzprime}.
The chiral spectrum of the models forms complete $E_6$ 
representations, whereas the additional vector--like 
multiplets may reside in incomplete multiplets. 
This is in fact an additional important property of the string, 
which affects compatibility with the gauge coupling data. 
The complete massless spectrum of the model was 
presented in ref. \cite{frzprime}. 
Spacetime vector bosons are obtained solely from the untwisted
sector and generate the observable and hidden gauge 
symmetries, given by:
\beqn
{\rm observable} ~: &~~~~~~~~SO(6)\times SO(4) \times 
U(1)_1 \times U(1)_2\times U(1)_3 \nonumber\\
{\rm hidden}     ~: &SO(4)^2\times SO(8)~.~~~~~~~~~~~~~~~~~~~~~~~\nonumber
\eeqn
The $E_6$ combination, 
\beq
U(1)_\zeta = U(1)_1+U(1)_2+U(1)_3 ~,
\label{u1zeta}
\eeq
is anomaly free whereas the orthogonal combinations of $U(1)_{1,2,3}$
are anomalous. The complete massless spectrum of the string model
and the charges under the gauge symmetries are given in ref. \cite{frzprime}.
Tables \ref{tableb} and \ref{tablehi}
show a glossary of the states in the model and 
their charges under the $SU(4)\times SO(4)\times U(1)_\zeta$
group factors, where we adopt the notation of ref. \cite{frdiphoton}.
The sextet states are in vector--like representations with respect
to the Standard Model, but are chiral under $U(1)_\zeta$. Thus, 
if $U(1)_\zeta$ is part of an unbroken $U(1)_{Z^\prime}$ combination
down to low scales, it protects the sextets, and corresponding bi--doublets, 
from acquiring a mass above the $U(1)_{Z^\prime}$ breaking scale.
The model also contains vector--like states that transform
under the hidden $SU(2)^4\times SO(8)$ group factors, with charges 
$Q_\zeta=\pm1$ or $Q_\zeta=0$. 

\begin{table}[!h]
\begin{tabular}{|c|c|c|c|}
\hline
Symbol& Fields in \cite{frzprime} & 
                         $SU(4)\times{SU(2)}_L\times{SU(2)}_R$&${U(1)}_{\zeta}$\\
\hline
$\FL$ &        $F_{1L},F_{2L},F_{3L}$&$\left({\bf4},{\bf2},{\bf1}\right)$&$+\oh$\\
$\FR$ &$F_{1R}$&$\left({\bf4},{\bf1},{\bf2}\right)$&$-\oh$\\
$\FbR$&${\bar F}_{1R},{\bar F}_{2R},{\bar F}_{3R},{\bar F}_{4R}$
                             &$\left({\bf\bar 4},{\bf1},{\bf2}\right)$&$+\oh$\\
$h$   &$h_1,h_2,h_3$&$\left({\bf1},{\bf2},{\bf2}\right)$&$-1$\\
$\Delta$&$ D_1,\dots,  D_7$&$\left({\bf6},{\bf1},{\bf1}\right)$&$-1$\\
$\bar\Delta$&$\Db_1,\Db_2,\Db_3,\Db_6$&$\left({\bf6},{\bf1},{\bf1}\right)$&$+1$\\
$S$&$\Phi_{12},\Phi_{13},\Phi_{23},\chi^+_1,\chi^+_2,\chi^+_3,\chi^+_5$
&$\left({\bf1},{\bf1},{\bf1}\right)$&$+2$\\
$\Sb$&$\bar\Phi_{12},\bar\Phi_{13},\bar\Phi_{23},\bar\chi^+_4$&$\left({\bf1},{\bf1},{\bf1}\right)$&$-2$\\
$\phi$&$\phi_1,\phi_2$&$\left({\bf1},{\bf1},{\bf1}\right)$&$+1$\\
$\phib$&$\bar\phi_1,\bar\phi_2$&$\left({\bf1},{\bf1},{\bf1}\right)$&$-1$\\
$\zeta$&$\Phi_{12}^-,\Phi_{13}^-,\Phi_{23}^-,\bar\Phi_{12}^-,\bar\Phi_{13}^-,\bar\Phi_{23}^-$&$\left({\bf1},{\bf1},{\bf1}\right)$&$\hphantom{+}0$\\
&$\chi_1^-,\chi_2^-,\chi_3^-,\bar\chi_4^-,\chi_5^-$&$$&$$\\
&$\zeta_i,\bar\zeta_i,i=1,\dots,9$&$$&$$\\
&$\Phi_i,i=1,\dots,6$&$$&$$\\
\hline
\end{tabular}
\caption{\label{tableb}
Observable sector field notation and associated states in \cite{frzprime}.}
\end{table}

\begin{table}[!h]
\begin{center}
\begin{tabular}{|c|c|c|c|}
\hline
Symbol& Fields in \cite{frzprime} & ${SU(2)}^4\times SO(8)$&${U(1)}_{\zeta}$\\
\hline
$H^+$&$H_{12}^3$&$\left({\bf2},{\bf2},{\bf1},{\bf1},{\bf1}\right)$&$+1$\\
&$H_{34}^2$&$\left({\bf1},{\bf1},{\bf2},{\bf2},{\bf1}\right)$&$+1$\\
$H^-$&$H_{12}^2$&$\left({\bf2},{\bf2},{\bf1},{\bf1},{\bf1}\right)$&$-1$\\
&$H_{34}^3$&$\left({\bf1},{\bf1},{\bf2},{\bf2},{\bf1}\right)$&$-1$\\
$H$&$H_{12}^1$&$\left({\bf2},{\bf2},{\bf1},{\bf1},{\bf1}\right)$&$0$\\
&$H_{13}^i,i=1,2,3$&$\left({\bf2},{\bf1},{\bf2},{\bf1},{\bf1}\right)$&$0$\\
&$H_{14}^i,i=1,2,3$&$\left({\bf2},{\bf1},{\bf1},{\bf2},{\bf1}\right)$&$0$\\
&$H_{23}^1$&$\left({\bf1},{\bf2},{\bf2},{\bf1},{\bf1}\right)$&$0$\\
&$H_{24}^1$&$\left({\bf1},{\bf2},{\bf1},{\bf2},{\bf1}\right)$&$0$\\
&$H_{34}^i,i=1,4,5$&$\left({\bf1},{\bf1},{\bf2},{\bf2},{\bf1}\right)$&$0$\\
$Z$&$Z_i,i=1,\dots,$&$\left({\bf1},{\bf1},{\bf8}\right)$&$0$\\
\hline
\end{tabular}
\end{center}
\caption{\label{tablehi}
Hidden sector field notation and associated states in \cite{frzprime}. }
\end{table}
As noted from table \ref{tableb} the string model contains the 
Higgs representations required to break the non--Abelian Pati--Salam
gauge symmetry \cite{patisalam}. These are ${\cal H}=F_R$ and 
$\bar{\cal H}$, being a linear combination of the four 
$\bar{F}_R$ fields. The decomposition of these fields under 
the Standard Model group is given by: 
\begin{align}
\bar{\cal H}({\bf\bar4},{\bf1},{\bf2})& \rightarrow u^c_H\left({\bf\bar3},
{\bf1},\frac 23\right)+d^c_H\left({\bf\bar 3},{\bf1},-\frac 13\right)+
                            {\bar {\cal N}}\left({\bf1},{\bf1},0\right)+
                             e^c_H\left({\bf1},{\bf1},-1\right)
                             \nonumber \\
{\cal H}\left({\bf4},{\bf1},{\bf2}\right) & 
\rightarrow  u_H\left({\bf3},{\bf1},-\frac 23\right)+
d_H\left({\bf3},{\bf1},\frac 13\right)+
              {\cal N}\left({\bf1},{\bf1},0\right)+ 
e_H\left({\bf1},{\bf1},1\right)\nonumber
\end{align}
The suppression of the left--handed neutrino masses favours
the breaking of the Pati--Salam (PS) gauge symmetry at the high scale
\cite{PSmodels, tnm}. The possibility of breaking the 
PS symmetry at a low scale was considered in refs. \cite{volkas,fg15}. 
Here we will take the PS breaking scale to be in the vicinity
of the string scale or slightly below. 
The VEVs of the heavy Higgs fields that 
break the PS gauge group leave an unbroken 
$U(1)_{Z^\prime}$ symmetry given by
\beq
U(1)_{{Z}^\prime} ~=~
{1\over {2}} U(1)_{B-L} -{2\over3} U(1)_{T_{3_R}} - {5\over3}U(1)_\zeta
~\notin~ SO(10),
\label{uzpwuzeta}
\eeq
that may remain unbroken down to low scales provided that $U(1)_\zeta$ is
anomaly free. 
Cancellation of the anomalies requires
that the additional vector--like quarks and leptons, 
that arise from the $10$ representation of $SO(10)$, 
as well as the $SO(10)$ singlet in the $27$ of $E_6$,
remain in the light spectrum.
The three right--handed neutrino states
are neutral under the low scale gauge symmetry 
and receive mass of the order of 
Pati--Salam breaking scale. 
The spectrum below the PS breaking
scale is displayed schematically 
in table \ref{table27rot}. 
The spectrum is taken to be supersymmetric down to the TeV scale.
As in the MSSM,
compatibility of gauge coupling unification with the
experimental data requires the existence of one vector--like 
pair of Higgs doublets, beyond the number of vector--like 
triplets. 
This is possible in the free fermionic
heterotic--string models due to the stringy doublet--triplet 
splitting mechanism \cite{dtsm}.
We allow also for the possibility of light states
that are neutral under the low scale gauge group.
In ref. \cite{frdiphoton} we showed that the string model 
contains all the ingredients to account for the LHC
di--photon excess, provided that the 
vector--like pairs of colour triplets and electroweak
doublets receive a mass of the order of the TeV
scale. This explanation is particularly 
appealing if the $U(1)_{Z^\prime}$ remains unbroken down 
to low scales. In this case the mass of the vector--like 
states can only be generated by the VEV of the $SO(10)$ singlets
$S_i$ and/or ${\phi_{1,2}}$ that breaks
the $U(1)_{Z^\prime}$ gauge symmetry. 
In this scenario the scale of the di--photon excess fixes 
the scale of the $U(1)_{Z^\prime}$ breaking to be of the 
order of the TeV scale. It is therefore of interest 
to examine the compatibility of this picture with 
the gauge coupling data.

\begin{table}[!h]
\noindent 
{\small
\begin{center}
{\tabulinesep=1.2mm
\begin{tabu}{|l|cc|c|c|c|}
\hline
Field &$\hphantom{\times}SU(3)_C$&$\times SU(2)_L $
&${U(1)}_{Y}$&${U(1)}_{Z^\prime}$  \\
\hline
$Q_L^i$&    $3$       &  $2$ &  $+\frac{1}{6}$   & $-\frac{2}{3}$   ~~  \\
$u_L^i$&    ${\bar3}$ &  $1$ &  $-\frac{2}{3}$   & $-\frac{2}{3}$   ~~  \\
$d_L^i$&    ${\bar3}$ &  $1$ &  $+\frac{1}{3}$   & $-\frac{4}{3}$  ~~  \\
$e_L^i$&    $1$       &  $1$ &  $+1          $   & $-\frac{2}{3}$  ~~  \\
$L_L^i$&    $1$       &  $2$ &  $-\frac{1}{2}$   & $-\frac{4}{3}$  ~~  \\
%
\hline
$D^i$       & $3$     & $1$ & $-\frac{1}{3}$     & $+\frac{4}{3}$  ~~    \\
${\bar D}^i$& ${\bar3}$ & $1$ &  $+\frac{1}{3}$  &   ~~$~2$  ~~    \\
$H^i$       & $1$       & $2$ &  $-\frac{1}{2}$   &  ~~$~2$ ~~    \\
${\bar H}^i$& $1$       & $2$ &  $+\frac{1}{2}$   &   $+\frac{4}{3}$   ~~  \\
\hline
$S^i$       & $1$       & $1$ &  ~~$0$  &  $-\frac{10}{3}$       ~~   \\
\hline
$h$         & $1$       & $2$ &  $-\frac{1}{2}$  &  $-\frac{4}{3}$  ~~    \\
${\bar h}$  & $1$       & $2$ &  $+\frac{1}{2}$  &  $+\frac{4}{3}$  ~~    \\
\hline
$\phi$       & $1$       & $1$ &  ~~$0$         & $-\frac{5}{3}$     ~~   \\
$\bar\phi$       & $1$       & $1$ &  ~~$0$     & $+\frac{5}{3}$     ~~   \\
\hline
%
$\zeta^i$       & $1$       & $1$ &  ~~$0$  &  ~~$0$       ~~   \\
\hline
\end{tabu}}
\end{center}
}
\caption{\label{table27rot}
\it
Spectrum and
$SU(3)_C\times SU(2)_L\times U(1)_{Y}\times U(1)_{{Z}^\prime}$ 
quantum numbers, with $i=1,2,3$ for the three light 
generations. The charges are displayed in the 
normalisation used in free fermionic 
heterotic--string models. }
\end{table}

\section{Gauge coupling analysis}\label{gcu}

In this section we analyse the 
compatibility of gauge coupling unification in the string inspired
model with the low energy gauge coupling data, 
where we may assume that the unification scale
is either at the GUT or string scales \cite{witten}. 
We examine the case in which the PS symmetry is broken at the 
string scale as well as the case in which is broken at 
an intermediate scale. 
We take the following values for the input parameters at the 
$Z$--mass scale \cite{pdg}:
\begin{align}
\bes
M_Z &= 91.1876\pm0.0021 \mbox{ GeV}\\
\alpha^{-1}&\equiv\alpha_{\mbox{\tiny{e.m.}}}^{-1}\left(M_Z\right)=127.944\pm0.014
\ees
&&
\bes
\left.\sin^2\theta_W\left(M_Z\right)\right|_{\overline{\mbox{\tiny{MS}}}}
&=0.23116\pm0.00012\\
\alpha_3\left(M_Z\right)&=0.1184\pm0.0007.
\ees
\label{pdg}
\end{align}
We also include the top quark mass of $M_t\sim 173.5$ GeV \cite{pdg}
and the Higgs boson mass of $M_H\sim 125$ GeV \cite{higgs} in our analysis.
String unification implies that the Standard Model gauge couplings are 
unified at the heterotic--string scale.
The one--loop renormalisation group equations (RGEs)
for the gauge couplings are given by
\beq 
\frac{1}{\alpha\left(M_X\right)}
=
\frac{1}{k_i\alpha_i\left(\mu\right)}
-\frac{b_i}{2\pi} \log\frac{M_X}{\mu^2}
+
\Delta_i^{\left(\mbox{\tiny{total}}\right)},
\label{rge}
\eeq
where $b_i$ are the one--loop beta--function coefficients, 
$\Delta_i^{\left(\mbox{\tiny{total}}\right)}$ represents
corrections two--loop and mixing effects, and $k_i=\{1,1,5/3\}$ 
for $i=3,2,1$. 
The analysis is most revealing at the one--loop level. 
Therefore, for the most part we limit our 
exposition to the one--loop investigation and 
give an estimate of the higher order corrections, 
which do not affect the overall picture. 
We obtain algebraic expressions for 
$\sin^2\theta_W\left(M_Z\right)$ 
and
$\alpha_3\left(M_Z\right)$
by solving the one--loop RGEs.
In our analysis, we initially assume the full spectrum of the $Z^\prime$ model 
between the unification scale, $M_X$,
and the $Z$--boson scale, $M_Z$, 
and treat all perturbations as effective threshold terms.
At the unification scale we have
\beq
\alpha_S\equiv\alpha_3(M_X)=\alpha_2(M_X)=k_1\alpha_Y(M_X),
\eeq
where $k_1={5}/{3}$ is the canonical $SO(10)$ normalisation.
We initially study the case in which the PS symmetry is broken
at the string scale. In this case the 
expression for 
$\left.\sin^2\theta_W\left(M_Z\right)\right|_{\overline{MS}}$
takes the general form
\begin{align}
\left.\sin^2\theta_W\left(M_Z\right)\right|_{\overline{MS}} = 
\Delta_{\mbox{\tiny {$Z^\prime$}}}^{\sin^2\theta_W}+
\Delta_{\mbox{\tiny {L.T.}}}^{\sin^2\theta_W}+
\Delta_{\mbox{\tiny {T.C.}}}^{\sin^2\theta_W}
\end{align}
with $\left.\alpha_3\left(M_Z\right)\right|_{\tiny{\overline{MS}}}$ 
having a similar form with corresponding $\Delta^{\alpha_3}$ corrections.
Here $\Delta_{Z^\prime}$ is the one--loop
contribution from the states of the $Z^\prime$ model
between the unification scale and the $Z$--boson mass scale. 
$\Delta_{\mbox{\tiny {L.T.}}}$
are corrections from the light thresholds, 
which consist of the light supersymmetric thresholds; 
the Higgs and the top mass thresholds; and the 
mass thresholds of the heavy vector--like matter states 
in the $Z^\prime$ model. 
The last term,
\beq
\Delta_{\mbox{\tiny {T.C.}}}^{\sin^2\theta_W} = 
\Delta_{\mbox{\tiny {Yuk.}}}^{\sin^2\theta_W}+
\Delta_{\mbox{\tiny {2-loop}}}^{\sin^2\theta_W}+
\Delta_{\mbox{\tiny {Conv.}}}^{\sin^2\theta_W},
\eeq 
includes the two--loop; kinetic mixing; Yukawa couplings and 
scheme conversion corrections. 
These corrections are found to be small and do not affect the overall picture. 
These effects can be absorbed into modifications of the light thresholds, 
which in any case are not fixed and can be varied. 
For $\sin^2\theta_W\left(M_Z\right)$ we obtain
\begin{align}
\bes \label{s2w}
\Delta_{\mbox{\tiny{$Z^\prime$}}}^{\sin^2\theta_W}
&=
\frac{3}{8}
+\frac{5\alpha}{16\pi} \left(b_2^{\mbox{\tiny{$Z^\prime$}}} -
                           b_1^{\mbox{\tiny{$Z^\prime$}}} 
                      \right)\log\frac{M_X}{M_Z};\\
\Delta_{\mbox{\tiny{L.T.}}}^{\sin^2\theta_W}&=
\frac{5\alpha}{16\pi}
\sum_i
           \left(b^{\mbox{\tiny{L.T.}}}_{1_i}- b^{\mbox{\tiny{L.T.}}}_{2_i}
                              \vphantom{\frac{1}{1}}
           \right) \log\frac{M_i}{M_Z},
\ees
\end{align}
where $M_i$ are the light mass thresholds and
$\alpha=\alpha_{\mbox{\tiny{e.m.}}}\left(M_Z\right)$. 
Similarly for $\alpha_3\left(M_Z\right)$, we have:
\begin{align}
\bes \label{a3}
&\Delta_{\mbox{\tiny{$Z^\prime$}}}^{\alpha_3}=
\frac{3}{8\alpha}+
\frac{1}{2\pi}
\left(\vphantom{\frac{1}{1}}
b_3^{\mbox{\tiny{$Z^\prime$}}} - \frac{3}{8} b_2^{\mbox{\tiny{$Z^\prime$}}} 
                         - \frac{5}{8} b_1^{\mbox{\tiny{$Z^\prime$}}} 
                      \right)\log\frac{M_S}{M_Z};\\
&\Delta_{\mbox{\tiny{L.T.}}}^{\alpha_3}=
\frac{1}{2\pi}
\sum_i \left(
\frac{5}{8} b^{\mbox{\tiny{L.T.}}}_{1_i} +
\frac{3}{8} b^{\mbox{\tiny{L.T.}}}_{2_i} - b^{\mbox{\tiny{L.T.}}}_{3_i}
                              \vphantom{\frac{1}{1}}
           \right) \log\frac{M_i}{M_Z}.
\ees
\end{align}

The predictions for gauge coupling observables at the $Z$--scale 
can therefore be seen to correspond to $0^{th}$ order predictions 
consisting of the first lines of eqs. (\ref{s2w}) and (\ref{a3})
plus the threshold corrections due to the decoupling of the 
different particles at their mass thresholds. The values of the 
beta function coefficients of these light thresholds are shown 
in table \ref{bfcoef}. The $0^{th}$ order coefficients are given by
\beqn
b_3^{\mbox{\tiny{$Z^\prime$}}}~~= & 
                        0 &   = ~~b_3^{\mbox{\tiny{MSSM}}}+ ~3,\nonumber\\
b_2^{\mbox{\tiny{$Z^\prime$}}}~~= & 
                        4 &   = ~~b_2^{\mbox{\tiny{MSSM}}}+ ~3,\nonumber\\
b_1^{\mbox{\tiny{$Z^\prime$}}}~~= & 
               \frac{48}{5} & =  ~~b_1^{\mbox{\tiny{MSSM}}} + 
                                                                 ~3.\nonumber
\eeqn
Hence, the $b_i^{\mbox{\tiny{$Z^\prime$}}}$ are identical to the 
$b_i^{\mbox{\tiny{MSSM}}}$ up to a common shift by 3, arising from the
vector--like colour triplets and electroweak doublets. As the 
$0^{th}$ order predictions for $\sin\theta(M_Z)$ and $\alpha_3(M_Z)$ 
only depend on the differences of the beta function coefficients,
the zeroes order predictions are identical to those that are
obtained in the MSSM.

\begin{table}[!ht]
\begin{center}
\begin{tabular}{|c||ccc|c|c|c|}
\hline
       $R$ & $b_{1}(R)$ & $b_2(R)$ & $b_3(R)$  & $ b_1-b_2$  &
       $\frac{5}{8}b_1+\frac{3}{8}b_2-b_3$ & factor\\
\hline
    ${\tilde g}$ & 0 &  0 & 2 & \,\,\,\,0 & $-2$ & $2\over 3$\\
    ${\tilde w}$ & 0 &  ${4\over3}$ & 0 & $-{4\over3}$ & \,\,\,\,${1\over2}$ & $2\over 3$\\
    ${\tilde\ell}_\ell$ & ${1\over{10}}$ & ${1\over6}$ & 0 & $-{1\over{15}}$ &\,\,  ${1\over8}$ & $1\over 3$\\
    ${\tilde\ell}_r$ & ${1\over{5}}$ & 0 & 0 & \,\,\,\,${1\over5}$ & \,\,\,\,${1\over8}$ & $1\over 3$\\
    ${\tilde Q}$ & ${1\over{30}}$ & ${1\over2}$ & ${1\over3}$ &$-{7\over{15}}$ 
                 & $-{1\over8}$ &$1\over 3$ \\
    ${\tilde d}_r$ & ${1\over{15}}$ &  0 & ${1\over6}$ &  \,\,\,\,${1\over{15}}$ &
           $-{1\over8}$ &$1\over 3$\\
    ${\tilde u}_r$ & ${4\over{15}}$ &  0 & ${1\over6}$ &  \,\,\,\,${4\over{15}}$ &
       \,\,\,   $0$ &$1\over 3$ \\
    ${\tilde h}$     & ${1\over{5}}$ &  ${1\over3}$ & 0 &
$-{2\over{15}}$ &
       \,\,\,   ${1\over{4}}$ & $2\over 3$\\
    $h$ & ${1\over{10}}$ &  ${1\over6}$ & 0 &
          $-{1\over{15}}$ & \,\,\,\,\,${1\over8}$ &$1\over 3$ \\
    $t$ & ${{17}\over{30}}$ & ${1\over2}$ & ${2\over3}$ & \,\,\,\,${1\over{15}}$ &
          $-{{1}\over{8}}$ &$2\over 3$\\
     $D+\tilde D$&$1\over 5$&$0$&$1\over 2$&\,\,\,\,$1\over 5$&$-{3\over 8}$&$1$\\
     $\bar D + \tilde {\bar D}$&$1\over 5$&$0$&$1\over 2$&\,\,\,\,$1\over 5$&$-{3\over 8}$&$1$\\
    $H+\tilde H$ &$3\over 10$&$1\over 2$&$0$&$-{1\over 5}$&\,\,\,\,${3\over 8}$&$1$\\
    $\bar H + \tilde {\bar H}$ &$3\over 10$&$1\over 2$&$0$&$-{1\over 5}$&\,\,\,\,${3\over 8}$&$1$\\     
\hline
\end{tabular}
\caption{
Beta function coefficients of the light thresholds in the string inspired 
$Z^\prime$ model. The factor in the last column indicates the spin
degeneracy factor.
}
\end{center}
\label{bfcoef}
\end{table}

The corrections due to the light thresholds are given by
\beqn
\delta\sin^2\big(\theta_W\big)_{ light} 
& =&\frac{5\alpha}{16\pi}\bigg(
   -{4\over 3} \log {M_{\tilde w} \over M_{Z}}
   -{1\over 5}\log {M_{{\tilde\ell}_\ell}\over M_{Z}}
   +{3\over 5}\log {M_{{\tilde\ell}_r}\over M_{Z}} 
   +{1\over 5}\log{M_{{\tilde d}_r}\over M_{Z}} \nonumber\\ 
& &-{7\over 5}\log {M_{{\tilde Q}_r}\over M_{Z}}
   +{4\over 5}\log {M_{{\tilde u}_r}\over M_{Z}} 
   -{4\over 15}\log {M_{\tilde h}\over M_{Z}}  
   -{2\over 15}\log {M_{h}\over M_{Z}}          \nonumber\\  
& &   +{1\over 15}\log {M_{t}\over M_{Z}}
   +{6\over 5}\log {M_{D}\over M_{Z}}
   -{6\over 5}\log {M_{H}\over M_{Z}}\bigg), ~~
                                            \label{s2wlight}\\
& & ~~\nonumber\\
\delta\big(\alpha_3^{-1}\big)_{ light} 
& =&\frac{1}{2\pi}\bigg(
        -2 \log {M_{\tilde g}\over M_{Z}}
        + {1\over 2}\log {M_{\tilde w}\over M_{Z}}
        -{3\over 8}\log {M_{{\tilde\ell}_\ell}\over M_{Z}}
        +{3\over 8}\log {M_{{\tilde\ell}_r}\over M_{Z}}  \nonumber  \\  
& &     -{3\over 8}\log {M_{{\tilde d}_r}\over M_{Z}}
        -{3\over 8}\log {M_{{\tilde Q}_r}\over M_{Z}}
        +{1\over 2}\log {M_{\tilde h}\over M_{Z}} 
        +{1\over 4}\log {M_{h}\over M_{Z}}\nonumber \\ 
& &     -{1\over 8}\log {M_{t}\over M_{Z}}
        -{9\over 4}\log {M_{D}\over M_{Z}}
        +{9\over 4}\log {M_{H}\over M_{Z}}\bigg) .
                                             \label{a3light}
\eeqn

\begin{center}
\begin{figure}[h]
  \rotatebox{90}{
\qquad \qquad\qquad \qquad\,\,\,\,{\Large$\alpha_{_S}(M_{Z})$}}
\includegraphics{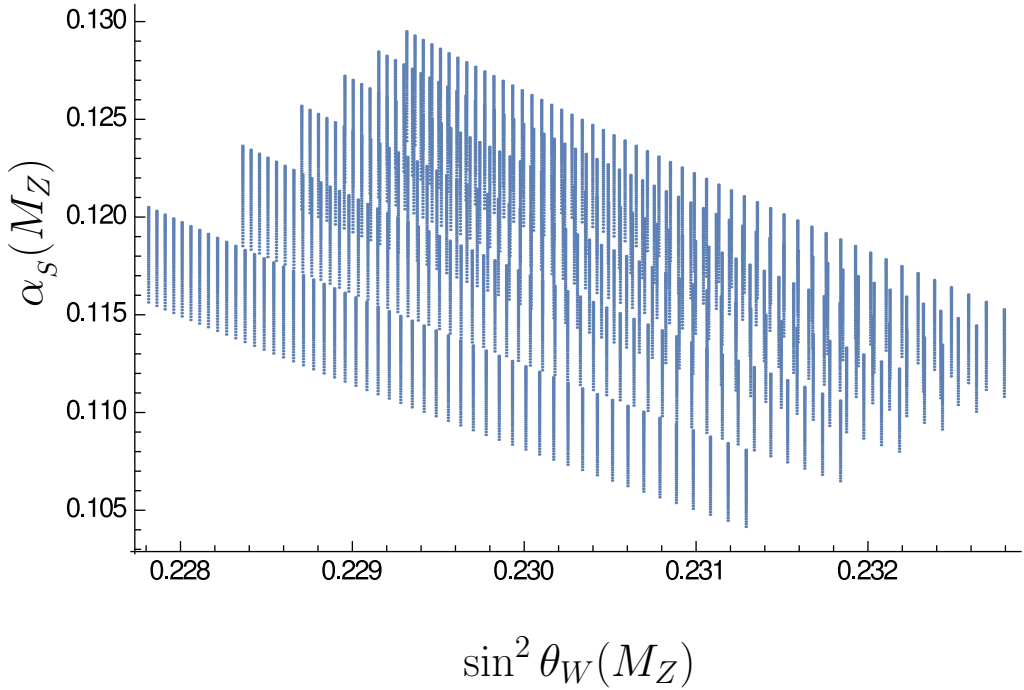}
 \rotatebox{90}{\qquad \qquad\qquad}{\Large$$\sin^{2}\theta_{W}(M_{Z})$$}
 \caption{\it Gauge coupling data at the electroweak scale in the presence
 of a light $Z^\prime$ and assuming unification at the heterotic--string scale.
 }
\label{plots2wa3}
\end{figure}
\end{center}

\begin{center}
\begin{figure}[h]
\includegraphics[width=15cm, angle=0]{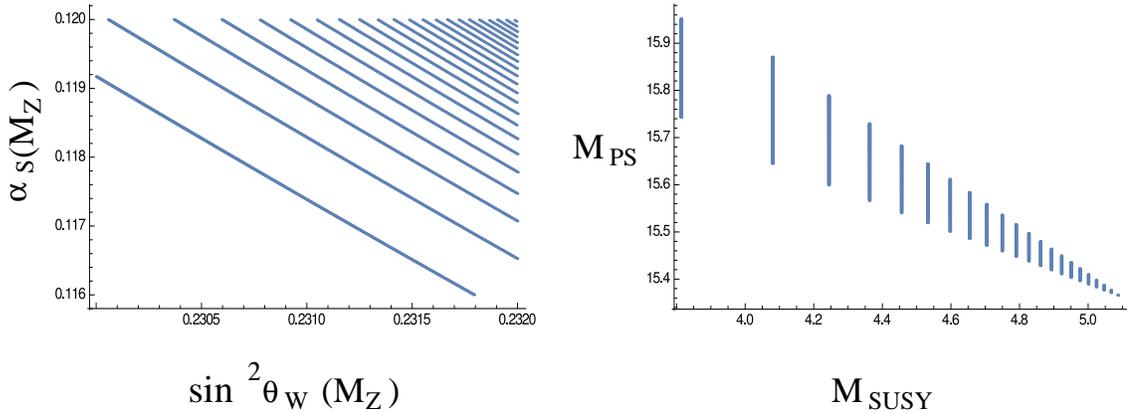}
 \caption{\it  The effect of intermediate Pati--Salam symmetry breaking scale 
in the $Z^\prime$ model pushes the supersymmetric thresholds beyond the 
LHC reach. 
The figure on the left displays the predictions for the gauge 
coupling parameters. The one on the right displays the PS scale versus a
common SUSY scale on a logarithmic scale $\log\left({M_{\rm PS}/M_Z}\right)$ vs 
$\log\left({M_{\rm SUSY}/M_Z}\right)$. }
\label{mpsvsmsusy}
\end{figure}
\end{center}

It is noted from eqs. (\ref{s2wlight}) and (\ref{a3light})
that if the vector--like colour triplets are degenerate in mass
with the vector--like electroweak doublets, then their threshold
corrections exactly cancel. In that case the predictions for 
$\sin^2\theta_W(M_Z)$ and $\alpha_3(M_Z)$ coincide exactly with those 
of the MSSM. The exact masses of these states depend of course on the 
details of their couplings to the $Z^\prime$ breaking VEV. 
Allowing for mass splitting of the order of a few TeV may be compensated 
by contributions from the supersymmetric states. Imposing the experimental 
limits on the supersymmetric particles and allowing for such mass differences
figure \ref{plots2wa3} shows a scatter plot of
$\sin^2\theta_W(M_Z)$ and $\alpha_3(M_Z)$, where the masses of the 
supersymmetric particles are varied independently. 
Next we study the predictions for the gauge coupling parameters 
with Pati--Salam breaking at an intermediate energy scale $M_{PS}$. 
The gauge symmetry is $SU(4)_C\times SU(2)_L\times SU(2)_R\times U(1)_\zeta$,
and $SU(3)_C\times SU(2)_L\times U(1)_Y\times U(1)_{Z^\prime}$,
above and below the intermediate Pati--Salam breaking scale, 
respectively. The weak hypercharge is given by\footnote{
$U(1)_C= 3 U(1)_{B-L}/2; ~U(1)_{\hat C}=  U(1)_C/\sqrt{3}$. } 
\beq
U(1)_Y~=~\frac{1}{3} U(1)_C + T_{3_R} \label{psu1y}
\eeq
with $k_C= 6$. When solving the RGEs for the low scale predictions we have to
distinguish the running above and below the intermediate breaking scale.
The RGEs and beta function coefficients below the symmetry breaking scale 
coincide with those of the $Z^\prime$ model discussed above. 
Above the symmetry breaking scale the spectrum differs from the 
standard Pati--Salam model due to the anomaly cancellation requirement 
of $U(1)_\zeta$. To ensure that $U(1)_\zeta$ is anomaly free, all 
the additional states above the intermediate breaking scale have to be
vector--like with respect to $U(1)_\zeta$. 
The Pati--Salam model contains an additional sextet field required for 
the missing--partner--like mechanism that gives heavy mass to the 
heavy Higgs states \cite{al}. Hence, anomaly cancellation with respect  
to $U(1)_\zeta$ demands another sextet in the spectrum with opposite 
$U(1)_\zeta$ charge. Similarly, the spectrum above the 
intermediate symmetry breaking scale contains two bi--doublet 
states with opposite $U(1)_{\zeta}$ charges, 
whereas only one pair of Higgs doublets remain below the 
intermediate scale. The beta function coefficients above the 
intermediate breaking scale are therefore 
\beq
b_4^{\mbox{\tiny{PS}}}~=~1~~~,~~~
b_2^{\mbox{\tiny{PS}}}~=~5~~~,~~~
b_{\mbox{\tiny{R}}}^{\mbox{\tiny{PS}}}~=~9~,
\eeq
which also takes into account the contribution of the heavy
Higgs states, and $b_2^{\mbox{\tiny{PS}}}, ~b_{\mbox{\tiny{R}}}^{\mbox{\tiny{PS}}}$
are the beta function coefficients of $SU(2)_L~,SU(2)_R$, respectively. 
The effect of the intermediate symmetry breaking scale is to add
correction terms to eqs. (\ref{s2w}) and (\ref{a3}), given by
\beqn
\Delta_{\mbox{\tiny{I.S.}}}^{\sin^2\theta_W}& = &
\frac{5\alpha}{16\pi}
  \left(b^{\mbox{\tiny{$Z^\prime$}}}_{1}- 
        \frac{3}{5}b^{\mbox{\tiny{PS}}}_{\mbox{\tiny{R}}}-
        \frac{2}{5}b^{\mbox{\tiny{PS}}}_{4}-
                   b^{\mbox{\tiny{$Z^\prime$}}}_{2}+
                   b^{\mbox{\tiny{PS}}}_{2}
                              \vphantom{\frac{1}{1}}
  \right) \log\frac{M_X}{M_{PS}},\label{PSs2w}\\
\Delta_{\mbox{\tiny{I.S.}}}^{\alpha_3} ~~~~& = &
\frac{1}{2\pi}
 \left( 
   \frac{3}{4} b^{\mbox{\tiny{PS}}}_{4}-
               b^{\mbox{\tiny{$Z^\prime$}}}_{3}-
   \frac{3}{8} b^{\mbox{\tiny{PS}}}_{\mbox{\tiny{R}}}+
   \frac{5}{8} b^{\mbox{\tiny{$Z^\prime$}}}_{1}+
   \frac{3}{8} b^{\mbox{\tiny{$Z^\prime$}}}_{2}-
   \frac{3}{8} b^{\mbox{\tiny{PS}}}_{2} 
    \vphantom{\frac{1}{1}}
 \right) \log\frac{M_X}{M_{PS}}.\label{PSa3}
\eeqn
Restricting to experimentally viable predictions for 
$\sin^2\theta_W(M_Z)$ and $\alpha_3(M_Z)$,
and varying $M_{PS}$ and a common SUSY breaking scale $M_{SUSY}$,
while keeping $M_X=1.1\times 10^{16}{\rm GeV}$ we obtain a relation between 
$M_{PS}$ and $M_{SUSY}$ which is displayed in figure \ref{mpsvsmsusy}. 
From the figure we note that reducing the intermediate Pati--Salam
symmetry breaking scale pushes the supersymmetric thresholds 
beyond the LHC reach. Nevertheless, the $Z^\prime$ breaking
scale remains at the TeV scale as the contribution of the 
extra vector--like colour triplets is canceled by that 
of the extra vector--like doublets.

\begin{center}
\begin{figure}[h]
\includegraphics[width=15cm, angle=0]{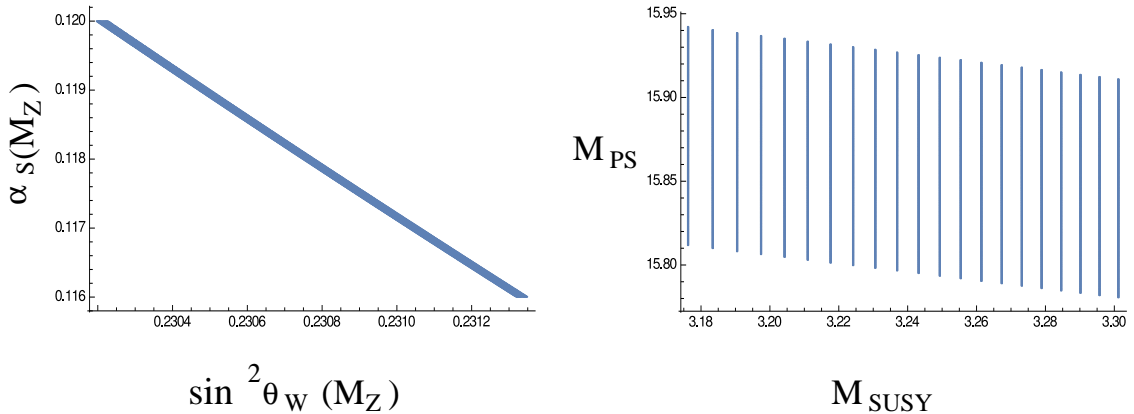}
 \caption{\it  The effect of additional heavy thresholds and an intermediate
symmetry breaking pushes the unification scale toward the perturbative
heterotic--string scale, while producing viable low scale predictions. 
The figure on the right displays the PS scale versus a
common SUSY scale on a logarithmic scale $\log\left({M_{\rm PS}/M_Z}\right)$ vs 
$\log\left({M_{\rm SUSY}/M_Z}\right)$. 
}
\label{mrvsmstring}
\end{figure}
\end{center}

The effects of the extra vector--like states above the Pati--Salam 
breaking scale may also mitigate the unification of the gauge coupling 
closer to the perturbative heterotic--string scale. Assuming
an additional pair of sextet fields, fixing $M_{SUSY}\sim 2{\rm TeV}$ and 
$M_X\sim 1\times 10^{17}{\rm GeV}$, we note that by varying the PS breaking
scale we obtain viable predictions for
$\sin^2\theta_W(M_Z)$ and $\alpha_3(M_Z)$.
These results are displayed in figure \ref{mrvsmstring}.

Split bi--doublet and sextet multiplets naturally 
appear in string models due to the stringy doublet--triplet stringy
mechanism, which depends on the assignment of boundary conditions 
in the basis vectors that break the $SO(10)$ symmetry to the 
Pati--Salam subgroup \cite{dtsm}. The model of
\cite{frzprime} contains three such pairs of untwisted sextets, 
and one additional pair from the twisted sectors, whereas there 
is no excess of vector--like bi--doublets. This is the 
case because the model of \cite{frzprime} utilises symmetric 
boundary conditions with respect to the internal manifold, 
whereas a model with asymmetric assignment would generate 
corresponding extra bi--doublets. The string models 
therefore contain all the ingredients to naturally produce
agreement with a di--photon excess as well as agreement with the
gauge coupling data at the electroweak scale. 

We may also consider the case of the left--right symmetric model
in which the $SO(10)$ symmetry is broken to $SU(3)\times U(1)_C\times 
SU(2)_L\times SU(2)_R$. We assume that $U(1)_\zeta$ 
charges admit the $E_6$ embedding. 
In this case the heavy Higgs states consists of the pair 
$
{\cal N}\left({\bf1},{\bf\frac{3}{2}},{\bf1},{\bf2},{\bf\frac{1}{2}}\right),~
{\bar{\cal N}}\left({\bf1},-{\bf\frac{3}{2}},{\bf1},{\bf2}, 
                                                   -{\bf\frac{1}{2}}\right).
$
The VEV along the electrically neutral component leaves unbroken
the Standard Model 
gauge group and the $U(1)_{Z^\prime}$ combination 
in eq. (\ref{uzpwuzeta}). 
We remark, however, that in the free fermionic LRS models \cite{lrs} 
the $U(1)_\zeta$ charges do not admit the $E_6$ embedding and 
we will argue in \cite{afr} that in a large class of string
models such construction is not possible. Here, we 
consider such models as purely field theory models
and study the effect on the low scale gauge coupling parameters. 
Above the symmetry breaking scale the spectrum coincides with that 
of table \ref{table27rot} with the right--handed fields 
arranged into doublet representations of $SU(2)_R$. Additionally, 
the spectrum contains the heavy Higgs states and a pair of 
Higgs bi--doublets with opposite $U(1)_\zeta$ charges. 
Crucially, here, the intermediate symmetry breaking does 
not require the existence of coloured states in the 
interval between $M_R$ and $M_X$, which may be incorporated 
in non--minimal extensions. Consequently, the beta
function coefficients above the intermediate 
symmetry breaking scale $M_R$ are
\beq
b_3^{\mbox{\tiny{R}}}~=~ 0~~~,~~~
b_2^{\mbox{\tiny{R}}}~=~ 5~~~,~~~
b_{\mbox{\tiny{R}}}^{\mbox{\tiny{R}}}~=~ 6~~~,~~~
b^{\mbox{\tiny{R}}}_{\mbox{\tiny{$\hat C$}}}~=~ 9~,~
\eeq
whereas the $b^{\mbox{\tiny{$Z^\prime$}}}_i$
below the intermediate breaking scale coincide with those given above. 
Here, $b_2^{\mbox{\tiny{R}}}$ is the beta function coefficient of 
$SU(2)_L$; $b_{\mbox{\tiny{R}}}^{\mbox{\tiny{R}}}$ is that of $SU(2)_R$; 
and $b^{\mbox{\tiny{R}}}_{\mbox{\tiny{$\hat C$}}}$ is that of the normalised 
$U(1)_C$ generator. 
The effect of the intermediate scale symmetry breaking is to add
correction terms for $\sin^2\theta_W(M_Z)$ and $\alpha_3(M_Z)$
given by
\beqn
\Delta_{\mbox{\tiny{I.S.}}}^{\sin^2\theta_W}& = &
\frac{5\alpha}{16\pi}
  \left(b^{\mbox{\tiny{$Z^\prime$}}}_{1}- 
        \frac{3}{5}b^{\mbox{\tiny{R}}}_{\mbox{\tiny{R}}}-
        \frac{2}{5}b^{\mbox{\tiny{R}}}_{\mbox{\tiny{$\hat C$}}}-
        b^{\mbox{\tiny{$Z^\prime$}}}_{2}+
        b^{\mbox{\tiny{R}}}_{2}
                              \vphantom{\frac{1}{1}}
  \right) \log\frac{M_X}{M_{R}},\label{ISs2w}\\
\Delta_{\mbox{\tiny{I.S.}}}^{\alpha_3} & = &
\frac{1}{2\pi}
 \left( 
   \frac{3}{8} 
    \left(
     b^{\mbox{\tiny{$Z^\prime$}}}_{2}-
     b^{\mbox{\tiny{R}}}_{2}-
     b^{\mbox{\tiny{R}}}_{\mbox{\tiny{R}}}- 
      \frac{2}{3} b^{\mbox{\tiny{R}}}_{\mbox{\tiny{$\hat C$}}}
    \right) +
\frac{5}{8} b^{\mbox{\tiny{$Z^\prime$}}}_{1} \vphantom{\frac{1}{1}}
    \right) \log\frac{M_X}{M_{R}}.\label{ISa3}
\eeqn

\begin{center}
\begin{figure}[h]
\includegraphics[width=15cm, angle=0]{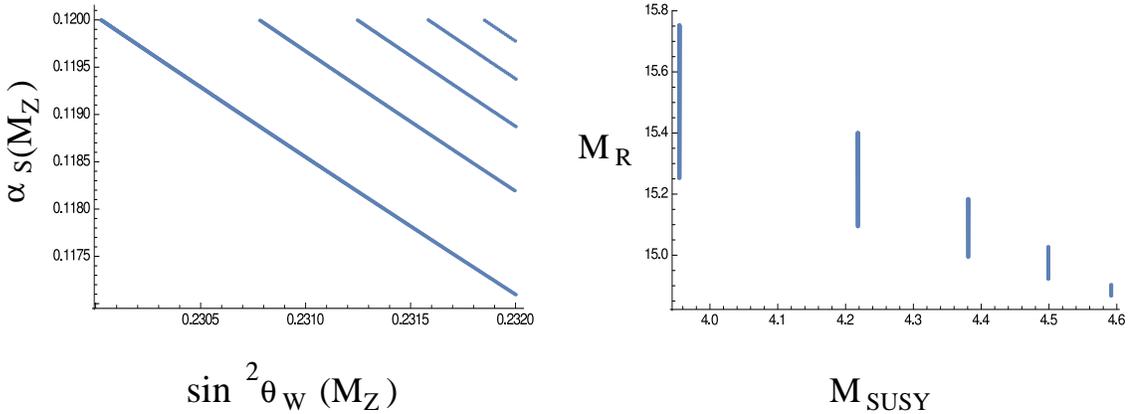}
 \caption{\it The effect of intermediate Left--Right Symmetry breaking scale 
in the $Z^\prime$ model pushes the supersymmetric thresholds beyond the LHC reach. 
The figure on the left displays the predictions for the gauge 
coupling parameters. The one on the right displays the LRS scale versus a
common SUSY scale, on a logarithmic scale
$\log\left({M_{\rm R}/M_Z}\right)$ vs 
$\log\left({M_{\rm SUSY}/M_Z}\right)$. }
\label{as2wlrs}
\end{figure}
\end{center}

As seen in figure \ref{as2wlrs} similar to the PS case the effect of the 
intermediate scale corrections to $\sin^2\theta_W(M_Z)$ and $\alpha_3(M_Z)$
is to shift the common SUSY threshold beyond the reach of the LHC. The 
figures should be viewed as illustrative, indicating the substantial impact that 
a low scale $Z^\prime$ may have on the anticipated signatures at
accessible energy scales. This should be contrasted with the
corresponding intermediate scale models \cite{dienes}, in  which the
impact of the intermediate scale corrections is milder.

\section{The di--photon events}\label{diphotons} 

In the low energy regime the superpotential \cite{frzprime} provides different 
interaction terms of the singlet fields $S_i$ and $\zeta_i$ which can be 
extracted 
from table ~\ref{table27rot}, among them we have
\beqn
\label{superpot}
\lambda^{ijk}_D S_i D_j \bar D_k + 
\lambda^{ijk}_H S_i H_j \bar H_k + 
\lambda^{ij}_h S_i H_j \bar h + 
\eta^i_{ \mathcal D} \zeta_i  {\mathcal D} \bar{ \mathcal D} +  
\eta^i_{ h} \zeta_i  h \bar{h} \,.
\eeqn
All these terms may comply with the di--photon excess reported by both the ATLAS 
and CMS experiments with a resonance around 750 GeV described by either the 
singlets $S_i$ or $\zeta_i$. Indeed, the presence of vector-like quarks, which 
is natural in heterotic-string models, facilitates the production of these 
states at the LHC.
In the following discussion we will consider the most simple and economic 
scenario in order to highlight the effects of the vector-like coloured states 
$D, \bar D$ and their role in the explanation of the di--photon excess.
 For this reason we assume that the resonance is reproduced by exchange of one 
of the singlet $S_i$ and we ignore the contribution of the $\zeta_i$ fields and 
of the coupling $S H \bar H$.
The real scalar component of one of the $S_i$ superfields acquires a VEV $v_S$ 
and breaks the extra $U(1)_{Z'}$ symmetry thus providing the mass of the $Z'$ 
gauge boson and of the $D, \bar D$ field through the coupling $\lambda_D$ in 
the superpotential (\ref{superpot}). Provided $v_S$ around the TeV scale, 
the mass of the singlet $S_i$, of the vector-like states $D, \bar D$ and of 
the $Z'$ lay in the TeV ballpark thus establishing a intimate relationship 
between the 750 GeV di--photon resonance and the presence of an additional 
spontaneously broken $U(1)_{Z'}$ gauge symmetry. Interestingly this can also 
be probed at the LHC in the lepto-production channel \cite{cfgprd, fg15}. 
Moreover, as we have already stated, in order to reproduce the di--photon 
excess it is enough to consider the impact of the vector-like coloured 
superfields $D, \bar D$ only. 
Therefore we assume $\lambda \equiv \lambda^{3ii}_D$ and we neglect all the other 
couplings. 
The fermionic components of $D_i$ and $\bar D_i$ can be rearranged 
into three Dirac spinors $\psi_{D_i}$, 
while the scalar components will provide six complex scalars 
$\tilde D_j$. The corresponding interaction Lagrangian can be parameterised as
\beqn
{\mathcal L} = - Y_D \, S \bar \psi_{D_i} \psi_{D_i} - \mu \, S |\tilde D_j|^2 \,,
\eeqn
where $S$ is the real scalar component of one of the $S_i$ singlet whose mass 
$M_S$ is identified with the 750 GeV resonance, $Y_D = \lambda/\sqrt{2}$ and 
$\mu$ is the corresponding soft-breaking term.
\begin{figure}
\centering
\subfigure[]{\includegraphics[scale=0.75]{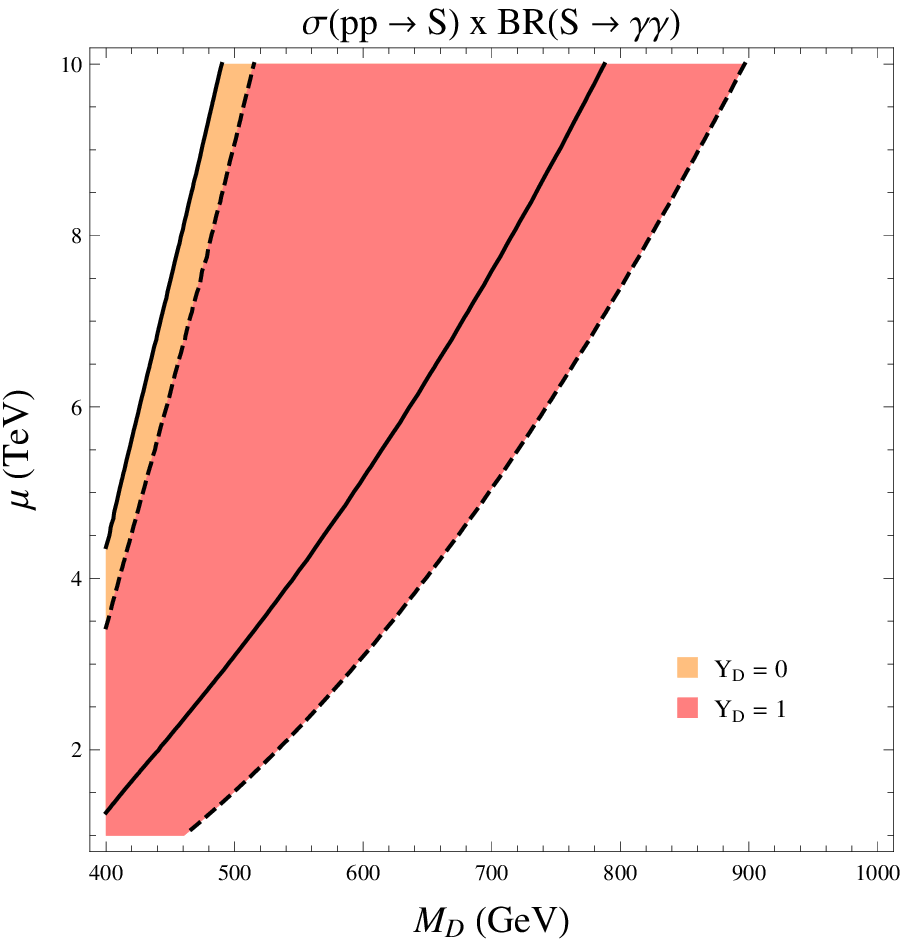}}
\subfigure[]{\includegraphics[scale=0.75]{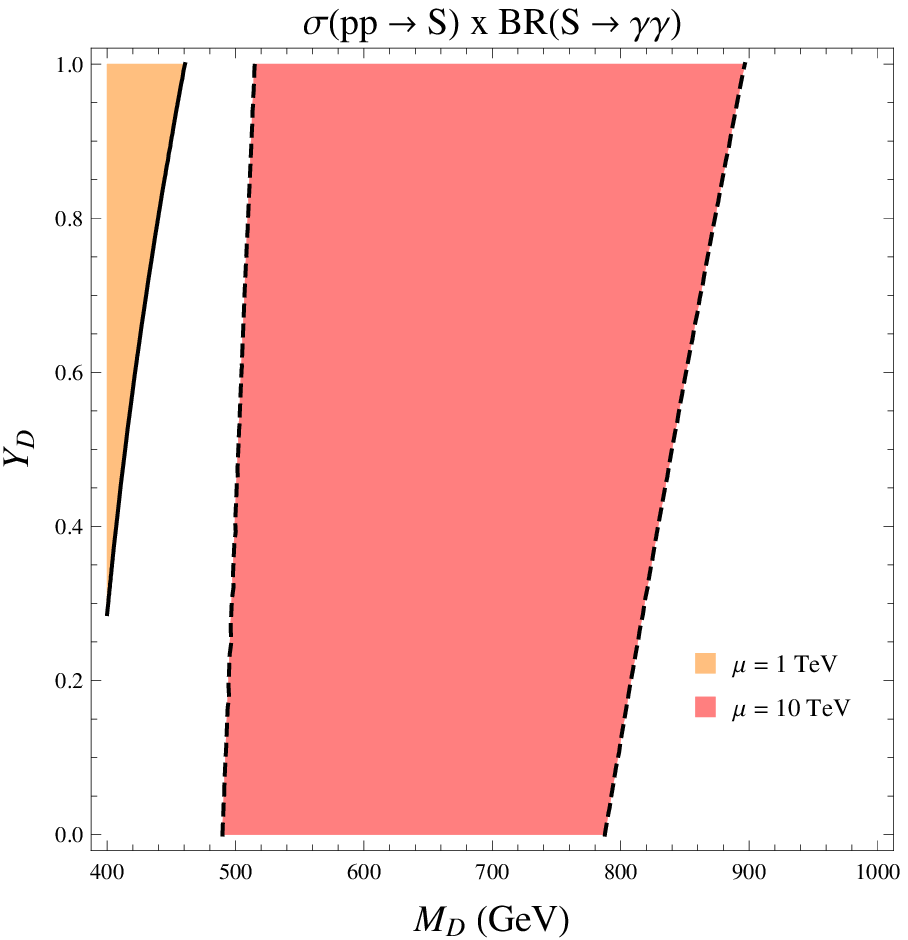}}
\caption{$\sigma(pp \rightarrow S) \times \rm{BR}(S \rightarrow \gamma \gamma)$ 
at 13 TeV LHC in (a) the $(M_D, \mu)$ plane for two values of the Yukawa 
coupling $Y_D$ and (b) in the $(M_D, Y_D)$ plane for two values of the scalar 
coupling $\mu$. The coloured regions correspond to a $2\sigma$ region of the 
measured cross section $4.5\pm1.9$ fb. \label{fig:xs}}
\end{figure}

\begin{figure}
\centering
\includegraphics[scale=0.75]{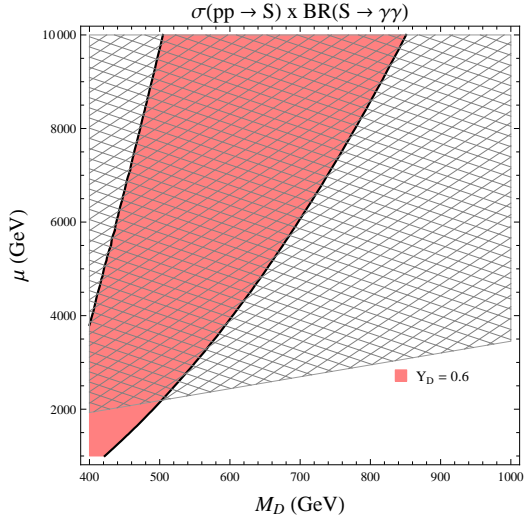}
\caption{$\sigma(pp \rightarrow S) \times \rm{BR}(S \rightarrow \gamma \gamma)$ 
at 13 TeV LHC in the $(M_D, \mu)$ plane for $Y_D = 0.6$. The coloured region corresponds to a $2\sigma$ interval around the 
measured cross section $4.5\pm1.9$ fb, while the hatched region is excluded by the requirement of colourless and electric neutral vacuum. \label{fig:xs2}}
\end{figure}

The LHC cross section of the di--photon production through the exchange of a 
scalar resonance in the $s$--channel is, in the narrow width approximation,
\beqn
\sigma (pp \rightarrow S \rightarrow \gamma \gamma) = 
\frac{1}{M_S \, s} C_{gg} \Gamma(S \rightarrow g g) \rm{Br}(S 
\rightarrow \gamma \gamma) 
\eeqn
where $M_S$ is the resonance mass, $C_{gg}$ the luminosity factor in the 
gluon--gluon channel and $\sqrt{s}$ the centre-of-mass energy. 
We assume that the main production mechanism occurs via gluon 
fusion with the corresponding luminosity factor at 13 TeV given by
\beqn
C_{gg} = \frac{\pi^8}{8} \int_{M_S^2/s}^1 \frac{dx}{x} g(x) g 
\left(\frac{M_S^2}{sx} \right) \simeq 2137 \,,
\eeqn
where $g(x)$ is the gluon distribution function and the value has been 
computed for $\sqrt{s} = 13$ TeV and for $M_S = 750$ GeV using MSTW2008NLO
\cite{mstw}. \\
The partial decay widths of $S$ into gluons and photons are
\beqn
\Gamma(S \rightarrow gg) &=& \frac{\alpha_S^2}{128 \pi^3} M_S^3 \bigg| 
\sum_f \frac{y_f}{m_f} A_{1/2}(\tau_f)  + \sum_s \frac{\mu_s}{2 m_s^2} 
A_0(\tau_s)  \bigg|^2 \,, \\
\Gamma(S \rightarrow \gamma \gamma) &=& 
\frac{\alpha^2}{256 \pi^3} M_S^3 \bigg| \sum_f N_c^f q_f^2 
\frac{y_f}{m_f} A_{1/2}(\tau_f)  + \sum_s N_c^s q_s^2 
\frac{\mu_s}{2 m_s^2} A_0(\tau_s)  \bigg|^2 \,,
\eeqn
where $m_f$ and $m_s$ are the masses of a generic fermion and scalar running in 
the loops, $y_f$ and $\mu_s$ the corresponding couplings to $S$ and $N_c$ the 
colour factor. As $D, ~\bar D$ are singlets of $SU(2)_L$, their electric charge 
$q$ coincides with the hypercharge $Y$.
The fermionic and scalar loop functions are given by
\beqn
A_{1/2}(\tau) = 2 [\tau + (\tau -1) f(\tau)]/\tau^2, \qquad 
A_{0}(\tau) = - [\tau - f(\tau)]/\tau^2
\eeqn
with $\tau_i = M_S^2/(4 m_i^2)$ and
\beqn
f(\tau) = \begin{cases} \arcsin^2 \sqrt{\tau}, & \mbox{if } \tau \le 1 \\ 
-\frac{1}{4} \left[ \log \frac{1 + \sqrt{1 - \tau^{-1}}}{1 - \sqrt{1- \tau^{-1}}} 
- i \, \pi\right]^2  & \mbox{if } \tau > 1 \end{cases} \,.
\eeqn
Assuming 
$\Gamma_{\rm tot} = \Gamma(S \rightarrow gg) + \Gamma(S \rightarrow \gamma \gamma)$, 
we show in figure \ref{fig:xs} the portion of the parameters space in which the 
di--photon excess can be reproduced in a $2\sigma$ region around the measured 
value $\sigma = 4.5\pm1.9$ fb reported by the ATLAS and CMS collaborations at 
13 TeV. For simplicity we assume 
$M_{\psi_{D_i}} \simeq M_{\tilde D_i} \simeq M_D$ 
and we present our results in the $(M_D, \mu)$ and $(M_D, Y_D)$ planes. 
The cross section is dominated by the complex scalar loops while the fermionic 
components of the supermultiplets $D, ~\bar D$ only provide a small contribution. 
Therefore, a huge Yukawa coupling is not strictly necessary as usually required 
in the literature, as its effect is compensated by a large soft--breaking term 
and relatively light squark--like states.  
Nevertheless, the di--photon cross section is also reproduced in regions of the parameter space 
characterised by big values of $Y_D$. Therefore, it is natural to ask if the running of the Yukawa coupling
up to the unification scale does not induce a loss of pertubativity at high energies. For this purpose
we have computed the corresponding $\beta$ function
\beqn
\beta_{Y_D} = Y_D \left( - \frac{4}{15} g_1^2  - \frac{19}{10} g_1'^2   -  \frac{16}{3} g_3^2    + 22 Y_D^2 + 
  2  \lambda_h^2 + 2  \lambda_H^2\right)
\eeqn
where, for the sake of simplicity, we have neglected the kinetic mixing 
and the tensor structure of the couplings. The contributions from the gauge sector, and in particular of the strong gauge group,
provide a decreasing evolution for $Y_D$ which could be prevented mainly by the $Y_D^3$ term.
This behaviour, due to the $SU(3)$ charge of the supermultiplets $D$ and $\bar D$,  
is similar to that of the top-quark in the SM in which the QCD corrections are responsible
for a monotonically decreasing $Y_t$ along all the RG running.
We have explicitly verified that $Y_D \sim 0.6$ still preserves its perturbativiy up to $10^{16}$ GeV.
The inclusion of the kinetic mixing would improve the perturbativity limit, even if only slightly. \\
For smaller values of the Yukawa coupling $Y_D$, the $D, ~\bar D$ scalar components running in the loops, which interact with the singlet $S$
through the the soft--breaking term $\mu$, represent the dominant contribution to the cross section. 
However, a large trilinear term may spoil the stability of the potential or induce a coloured and electric charged vacuum 
(see for instance \cite{gu} for studies related to the 750 GeV excess).
Preventing this situation will introduce an upper bound on the $\mu$ term whose exact value obviously depends on the details of the soft--breaking Lagrangian.
This would clearly require a dedicated study of the parameter space, here we give some comments. 
The relevant part of the scalar potential can be parameterised in the following form
\beqn
V(S,  D, \bar D) &=& m_{D_i}^2 |D_i |^2 + m_{\bar D_i}^2 |\bar D_i |^2 + \frac{1}{2} M_S^2 S^2  + \frac{\mu_S}{3} S^3 + \frac{\lambda_1}{4} S^4 + \mu \, S (|D_i |^2 + |\bar D_i |^2 ) \nonumber \\
&+&  \frac{ \lambda_2}{2} S^2  |D_i |^2 + \frac{\lambda'_2}{2} S^2 |\bar D_i |^2   + \lambda_3 ( D_i \bar D_i ) ( D_i^\dag \bar D_i^\dag )  \nonumber \\
&+& \lambda_4 \left(    ( D_i^\dag  D_j ) ( D_j^\dag  D_i )   +    ( \bar D_i^\dag  \bar D_j ) ( \bar D_j^\dag  \bar D_i )   + 2 ( D_i \bar D_j ) ( D_j^\dag \bar D_i^\dag )  \right) \nonumber \\ 
&+&   \lambda_5 |D_i |^2 |D_j |^2  +  \lambda'_5 |\bar D_i |^2 |\bar D_j |^2    + \lambda_6     |D_i |^2 |\bar D_j |^2 ,
\eeqn
where $S$ is the physical real scalar component, $i,j = 1,2,3$ and the quartic couplings have been extracted from the $F$ and $D$ terms
\beqn
&& \lambda_1 = \frac{1}{8} g_1'^2 Q^{' 2}_S,  \qquad  \lambda_2 =  \lambda_D^2  +  g_1'^2 Q^{' }_S Q^{' }_{D}, \qquad \lambda'_2 = \lambda_D^2  +  g_1'^2 Q^{' }_S Q^{' }_{\bar D} \\
&&\lambda_3 = \lambda_D^2, \qquad \lambda_4 = \frac{1}{4} g_3^2 ,\qquad \lambda_5 = - \frac{1}{12} g_3^2 + \frac{1}{18} g_1^2  + \frac{1}{2} g_1'^2 Q^{' 2}_{D}   \\
&&\lambda'_5 = - \frac{1}{12} g_3^2 + \frac{1}{18} g_1^2  + \frac{1}{2} g_1'^2 Q^{' 2}_{\bar D} , \qquad \lambda_6 = - \frac{1}{6} g_3^2  - \frac{1}{9} g_1^2 +  g_1'^2 Q^{' }_{D} Q^{'}_{\bar D} ,
\eeqn
with $Q'$ being the charge under the $U(1)_{Z'}$ gauge group.
We require $\langle S \rangle = 0$ (notice that in the parameterisation of the scalar potential given above, 
the scalar singlet has already undergone spontaneous symmetry breaking) and $\langle D_i \rangle = \langle \bar D_i \rangle = 0$, 
thus identifying the region of the parameter space in which the occurrence of a coloured vacuum is avoided. 
To simplify the discussion we study the scenario of a flavour independent vacuum, namely $v_D \equiv \langle D_i \rangle =  \langle \bar D_i^\dag \rangle$. 
In this case the minimisation conditions read
\beqn
&& 6 \mu |v_D|^2 +  ( M_S^2 + 6 \alpha |v_D|^2 ) v_S + \mu_S v_S^2 + \lambda_1 v_S^3 = 0, \nonumber \\
&& |v_D| (2 M_D^2 + 6 \beta |v_D|^2 +  \alpha \, v_S^2 + 2 \mu \, v_S) = 0 .
\eeqn
with $\alpha = \lambda_2/2 + \lambda'_2/2$ and $\beta = \lambda_3 + 4\lambda_4 + \lambda_5 + \lambda'_5 + \lambda_6$.
In general, the destabilising effect of a large $\mu$ term can be counterweighted by large quartic couplings. 
In this scenario the latter are mainly controlled by $Y_D = \lambda_D/\sqrt{2}$ and the strong coupling constant $g_3$. 
We show in figure \ref{fig:xs2} the $2\sigma$ band around the central value of the di--photon cross section for $Y_D = 0.6$
and the corresponding excluded region in the $(M_D, \mu)$ plane. 
The bound is quite restrictive allowing, in this simplified setup, for a parameter space with $\mu \lesssim 2$ TeV and $M_D \lesssim 500$ GeV. 
We stress again that this analysis is far from being exhaustive, 
while its only purpose is to show how the di--photon excess can be naturally 
accommodated in heterotic--string scenarios where the $U(1)_{Z'}$ 
gauge symmetry is broken around the TeV scale. We have neglected, for instance, 
the impact of the $S H \bar H$ interaction which would increase, in general, 
the partial decay width into photons and thus broaden the preferred parameter 
space. 
As a side effect this would relax the necessity of a either large Yukawa coupling or soft--breaking term
and it will also provide more involved decay patterns through the mixing with the 
$H$ and $\bar H$ fields.

\section{The impact of the $D$-terms}
\begin{figure}[h]
\centering
\includegraphics[scale=0.6]{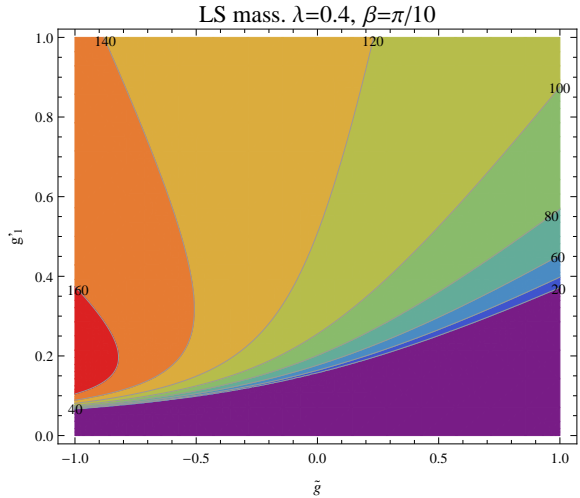}
\caption{Contour plot of lightest scalar eigenvalue of matrix (\ref{CPEven}). 
$v_S = 2.5$ TeV $M_{A_{\lambda}} = 500$ GeV. \label{LS}}
\end{figure}

The presence of an extra abelian factor together with the dynamical generation 
of a $\mu$-term supply our model with the minimal set of tools to 
relieve the tree-level MSSM hierarchy between the $Z$ and Higgs masses. 
To explore the low-energy scalar spectrum that can be naturally covered 
by the parameter space, we focus on the simple scenario involving only the 
fields interacting through the coupling $\lambda^{ijk}_H$ in (\ref{superpot}).
The neutral scalar components will then include 9 supermultiplets; 6 from 
$H, \bar{H}$ plus other 3 from the SM singlet $S$. 
Among different possible settings a viable one is achievable from 
\beqn \label{set1}
\langle H_{1,2} \rangle\, =\, \langle \bar{H}_{1,2} \rangle \,= \,\langle 
S_{1,2} \rangle \,=\, 0,
\eeqn
with non--zero VEVs concerning only the third generation
\beqn \label{set2}
\langle H_3\rangle = \frac{1}{\sqrt{2}} \left(\begin{array}{c} v_d \\
0 \end{array}\right),  \qquad 
\langle \bar{H}_3\rangle = \frac{1}{\sqrt{2}} \left(\begin{array}{c} 0 \\
v_u \end{array}\right) , \qquad 
\langle S_3\rangle = \frac{v_S}{\sqrt{2}} , 
\eeqn
where $v_u = v \sin \beta$ and $v_d = v \cos \beta$.
The setting in (\ref{set1}-\ref{set2}) is not the only one capable to 
minimise the scalar potential and break the symmetry down to 
$SU(3)\times U(1)_{em}$. 
It is nevertheless the one with the simplest and more MSSM-like structure. 
Given the illustrative purpose of this section, 
we take
$\lambda^{ijk}_H$ and the soft-SUSY masses to be flavour-diagonal
and real parameters. 
The part of the potential relevant to the spontaneous breaking analysis
contains only the (scalar component of the) fields $H_3, \bar{H}_3,$ and $S_3$
\beqn \label{pot}
V_{Higgs} &=& \tilde{m}^2_H |H|^2 + \tilde{m}^2_{\bar{H}} |\bar{H}|^2 + 
\tilde{m}^2_S |S|^2 - \left(\lambda_H \,A_{\lambda} H\,\bar{H}\,S + 
\text{h.c.} \right) \nonumber \\
&+& \lambda_H^2 \left(|H \bar{H}|^2 + |H|^2|S|^2 + 
|\bar{H}|^2|S|^2 \right) \nonumber \\
&+& \frac{1}{2}\,g_2^2 \left(H^{\dagger}\frac{\sigma^{\mu}}{2}H + 
\bar{H}^{\dagger}\frac{\sigma^{\mu}}{2}\bar{H}\right)^2 
+ \frac{1}{2} g_1^2 \left(\frac{1}{2}|\bar{H}|^2 - \frac{1}{2}|H|^2   
\right)^2 \nonumber \\
&+& \frac{1}{2} {g'_1}^2 \left(Q'_{\bar{H}}|\bar{H}|^2 + Q'_{H}|H|^2 + 
Q'_{S}|S|^2  \right)^2 \, ,
\eeqn
with the generator of the extra Abelian group given in the form which 
includes the mixing $g'_1\,Q'_{f} = g'_1 Y_{f}' + \tilde{g} Y_{f}$, where 
$Y_{f}'$ and $Y_{f}$ are, respectively, the charges under $U(1)_{Z'}$ and 
$U(1)_Y$.
As customary, the trilinear (dimensionful) coefficient has been written in the 
form $\lambda_H\,A_{\lambda}$. 
The three soft-masses 
$\tilde{m}_{H\, 3,3}^2, \tilde{m}_{\bar{H}\,3,3}^2, \tilde{m}_{S\,3,3}^2$ 
non--trivially solve the tadpole--conditions to accommodate for the
VEVs structure of (\ref{set1}-\ref{set2}). Putting such values in the 
neutral-boson mass matrices and considering the large $v_S$ limit we obtain
\beqn
m^2_{Z} = \frac{v^2}{4} \left(g_1^2 + g_2^2\right)\, ,\,\, m^2_{Z'} = 
\left(Q'_{S}\, g'_1\, v_S\right)^2 = \left(Y'_{S}\,g'_1\,v_S\right)^2 \, . 
\eeqn
By requiring 
\beqn \label{set3}
\tilde{m}_{H\,1,1}^2 = 
\tilde{m}_{H\,2,2}^2 \,\, , \tilde{m}_{\bar{H}\,1,1}^2 = 
\tilde{m}_{\bar{H}\,2,2}^2\,\, , \tilde{m}_{S\,1,1}^2 = 
\tilde{m}_{S\, 2,2}^2\, ,
\eeqn
the $9\times9$ CP-odd mass matrix can be analytically diagonalised. 
In the Landau gauge the two massless Goldstone bosons are promptly found and 
the remaining 7 masses are a degenerate ensemble of the independent set:
\beqn
\left(m^2_{1}, m^2_{2}, m^2_{3},m^2_{A_{\lambda}} \right)\, .
\eeqn
The eigenvalues $m^2_{1-3}$ are uniquely linked to the three independent 
soft masses of (\ref{set3}) and consequently are all double degenerate. 
The eigenvalue dubbed as $m^2_{A_{\lambda}}$ is connected to the trilinear 
soft term. In the limit of large $v_S$ we find
\beqn \label{cpodd_ma}
m^2_{A_{\lambda}} = 
\sqrt{2}\, v_S\, \lambda_H \frac{A_{\lambda}}{\sin(2\beta)} \,,
\eeqn
where $\tan \beta$ $= v_u / v_d $. The correspondence with the MSSM is
clear once we identify the effective $\mu$-term 
$\mu_{eff} = v_S\,\lambda_H\,/\sqrt{2}$.
All the soft-masses in (\ref{pot}) can thus be traded for the CP-odd 
eigenvalues and, via tadpole conditions, for the non-zero VEVs. 
The mass matrix for the charged Higgs scalars\footnote{We are always 
considering only the supermultiplets $H$, $\bar{H}$ and $S$.} can 
similarly be analytically
diagonalised. The eigenvalues are simply linked to the $W$ mass and 
the CP-odd masses. In the Landau gauge we find one
massless Goldstone while the remaining independent masses are given by 
(for $v_S \gg v$)
\beqn
&&\left(m^2_{1} + M_W^2\,\cos(2\beta), \,m^2_{2} - 
M_W^2\,\cos(2\beta), m^2_{A_{\lambda}} + M_W^2 - 
\frac{\lambda^2 v^2}{2} \right),
\eeqn
with degeneracy inherited from the CP-odd structure. 
The CP-even mass matrix is mostly diagonal with mixing involving 
only the third generations of $H, \bar{H}$, and $S$.
The 6 eigenvalues in the diagonal are degenerate to the 
corresponding CP-odd partners $m^2_{i= 1,2,3}$.
The remaining $3\times3$ block to be diagonalised includes the matrix elements
\beqn \label{CPEven}
m^2_{1,1} &=& M_Z^2\, \cos^2 \beta + 
4\,M_Z^2\,\left(\frac{g'_1\,Q'_{H}}{\bar{g}}\right)^2\,\cos^2 \beta + 
\Delta\,\,\sin^2 \beta , \nonumber \\
m^2_{2,2} &=& M_Z^2\, \sin^2 \beta + 
4\,M_Z^2\,\left(\frac{g'_1\,Q'_{\bar{H}}}{\bar{g}}\right)^2\,\sin^2 \beta + 
\Delta\,\,\cos^2 \beta , \nonumber \\
m^2_{3,3} &=& {M'_Z}^2  + 
\Delta \left(\frac{M_Z\,\sin (2\beta)}{\bar{g}\,v_S}\right)^2 , \nonumber \\
&& \nonumber \\
m^2_{1,2} &=& \, \cos \beta \,\sin \beta \, \left(- M^2_Z - 
\Delta + \frac{4\,M_Z^2}{\bar{g}^2}\,\left( \lambda^2 + 
{g'_1}^2\,Q'_{H}\,Q'_{\bar{H}}  \right)\right)\, , \nonumber \\
m^2_{1,3} &=& \,\cos \beta \left(2\,\frac{M_Z\,v_S}{\bar{g}}\right)\,\left(- 
\frac{\Delta}{v_S^2}\,\sin^2 \beta + \lambda^2 + 
{g'_1}^2\,Q'_{H}\,Q'_{S} \right) \, ,\nonumber \\
m^2_{2,3} &=& \,\sin \beta \left(2\,\frac{M_Z\,v_S}{\bar{g}}\right)\,\left(- 
\frac{\Delta}{v_S^2}\,\cos^2 \beta + \lambda^2 + {g'_1}^2\,Q'_{S}\,Q'_{\bar{H}} \right)
\eeqn
where 
\beqn
\bar{g}^2 &=& g_1^2 + g_2^2 \,,\,\,\,\,\,\,\,
\Delta = \frac{\bar{g}^2\, M^2_{A_{\lambda}}\,v^2_S}{M_Z^2\, \sin^2 (2 \beta)+
\bar{g}^2\,v_S^2}\, .
\eeqn
The numerical diagonalisation of the previous mass matrices easily reveals 
large branches of the parameter space with tree-level eigenvalues that
elude the MSSM hierarchy between the lightest scalar (LS) and $M_Z$ 
(Fig~\ref{LS}). 
To obtain an analytical estimation of the impact of the $D$--terms 
we minimise the expectation value of the CP-even mass matrix with the 
vector $(\cos \beta, \sin \beta,0)$
\cite{Ellwanger:2009dp}. 
The result represents an upper limit for its smallest eigenvalue
\beqn \label{hier}
M^2_{h} \leq M^2_Z\,\cos^2(2\,\beta) + 
\frac{v^2}{2}\,\lambda^2\,\sin^2(2\,\beta) + 
{g'_1}^2\,v^2\,\left(Q'_{H}\,\cos^2 \beta + 
Q'_{\bar{H}}\,\sin^2 \beta \right)^2   .
\eeqn
\begin{figure}
\centering
\subfigure[]{\includegraphics[scale=0.5]{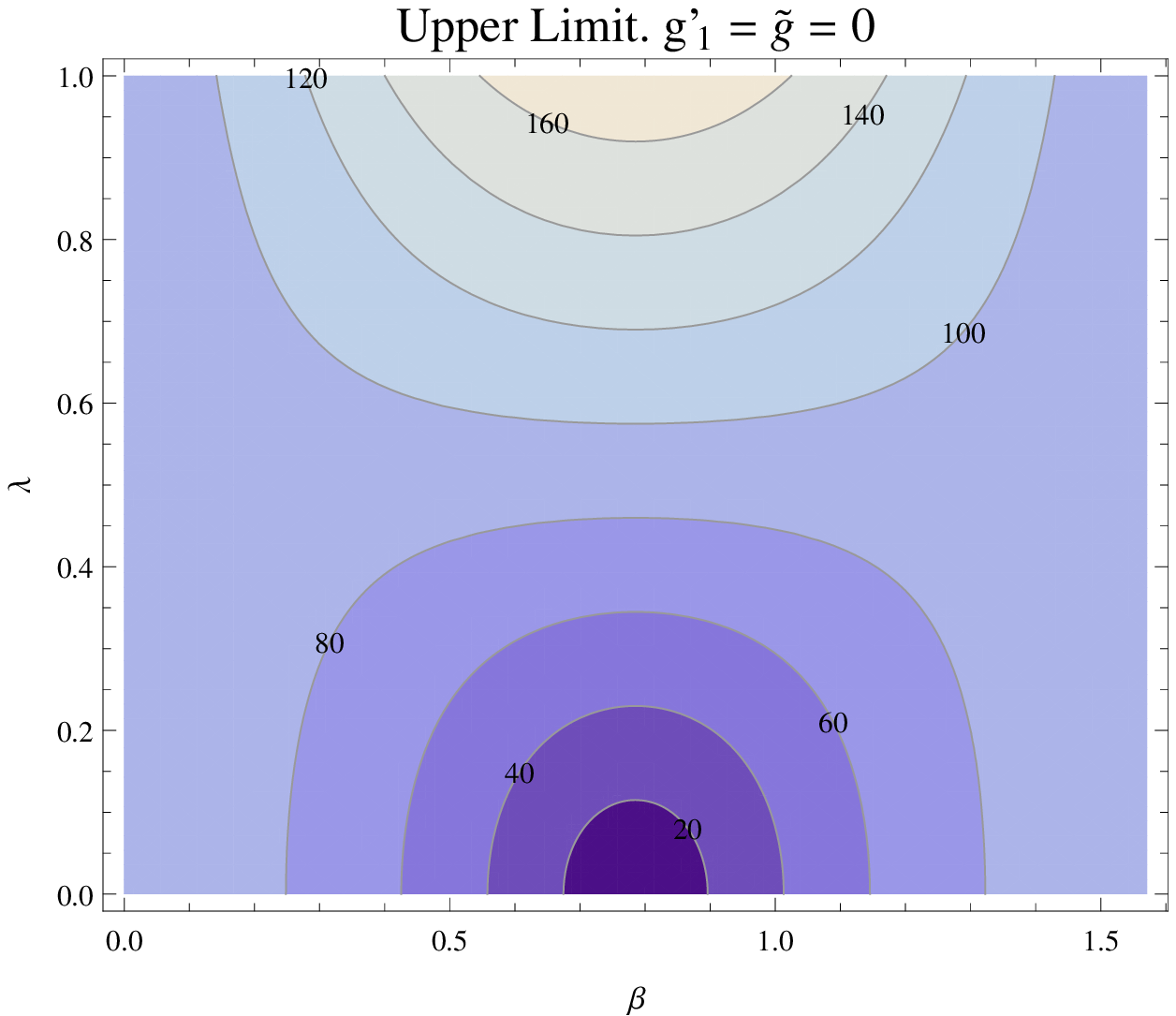}}
\subfigure[]{\includegraphics[scale=0.5]{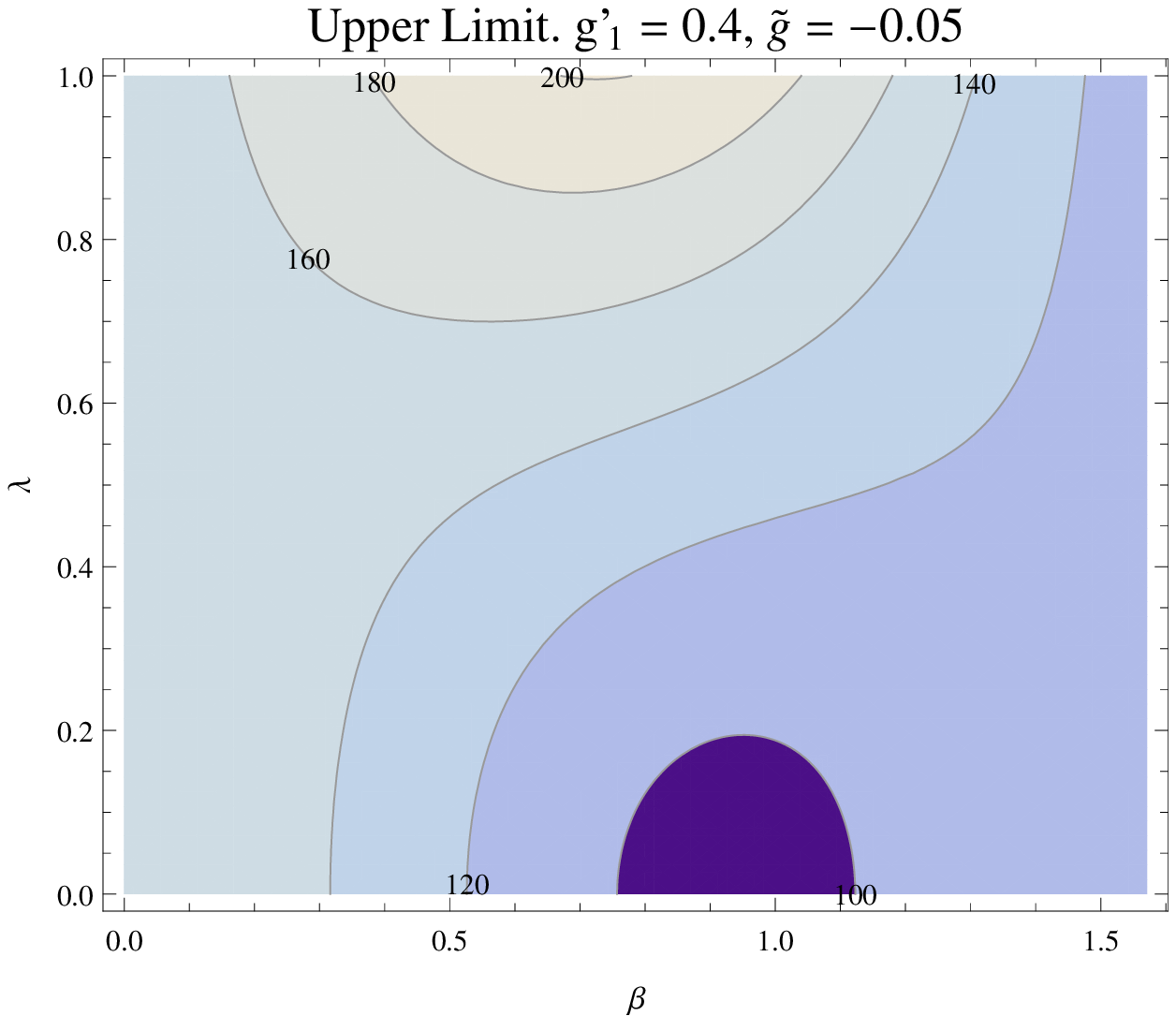}}
\caption{Contour plot of upper bounds for LS mass. \label{Upper}}
\end{figure}
In the formal limit $g'_1,\tilde{g} \rightarrow 0 $ 
we recover the upper bound of the NMSSM 
\cite{Quiros:1998bz}-\cite{Ellwanger:2009dp} 
and a further limit, $\lambda_H \rightarrow 0$, we obtain
the MSSM one. As known, the singlet extension of the MSSM is a first step to 
increase the tree-level value of the LS. 
The positive contribution of the $U(1)_{Z'}$-related $D$--terms
in (\ref{hier}) allows even larger upper bounds (Figs.~\ref{Upper}).

\section{Conclusions}

The Standard Model of particle physics continues to reign supreme in providing 
viable parameterisation for subatomic observational data. Incorporating 
gravitational phenomena mandates the extension of the Standard Model, 
with string theory providing: 
\begin{itemize}
\item
minimal departure from the point particle hypothesis underlying the Standard Model. 
\item
mathematically self--consistent framework for perturbative quantum gravity. 
\item
mathematically self--consistent framework to develop a phenomenological 
approach to explore the synthesis of the gauge and gravitational interactions. 
\end{itemize}
Phenomenological string models 
constructed in the so called fermionic formulation 
\cite{fsu5,slm,PSmodels, lrs, su62} correspond to $Z_2\times Z_2$ orbifolds at
enhanced symmetry points in the toroidal moduli space \cite{z2xz2}.
These models reproduce the main characteristic of the Standard Model 
spectrum, {\it i.e.}
the existence of three chiral generations and their embedding 
in spinorial 16 representations of $SO(10)$. 

\begin{figure}[!h]
\begin{center}
\includegraphics[width=10cm]{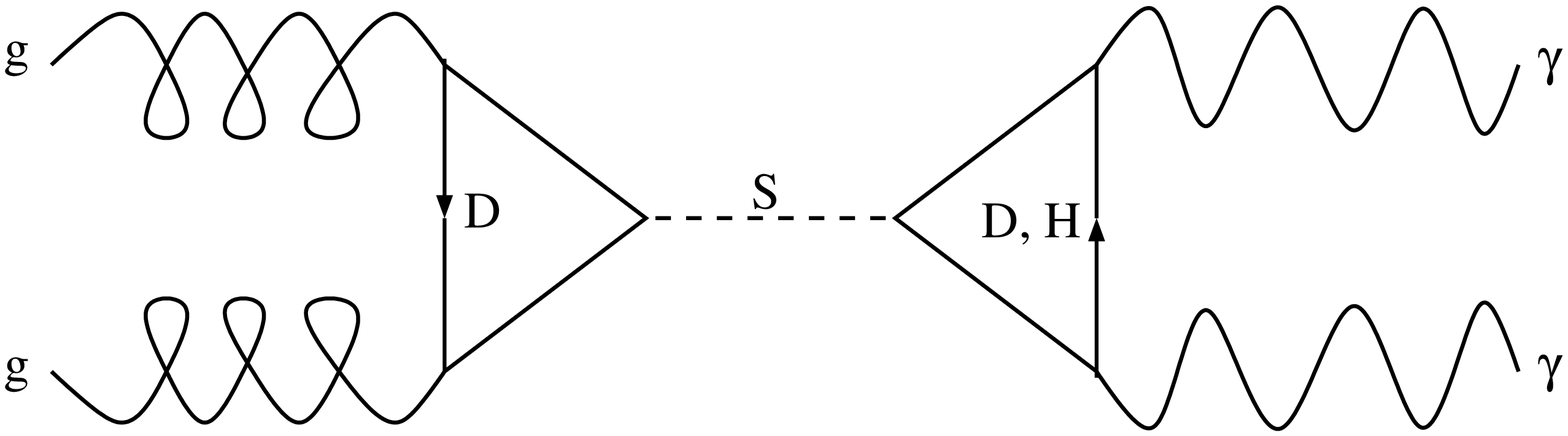}
\caption{Production and di--photon decay of the Standard Model 
singlet scalar state.}
\label{didiagram}
\end{center}
\end{figure}

Indications for di--photon excess at the LHC 
will provide a vital clue in seeking the fundamental origins of the 
Standard Model.
Such excess, and absence of any other observed signatures, 
is well explained
as a resonance of a Standard Model singlet scalar field, 
which is produced and decays via triangular loops 
incorporating heavy vector--like states as
depicted schematically in figure \ref{didiagram}. 
All the ingredients for producing the diagram depicted in figure
\ref{didiagram} arise naturally in the string derived $Z^\prime$ model
\cite{frzprime, frdiphoton}. 
The chirality of the Standard Model singlet and the vector--like 
states under $U(1)_{Z^\prime}$ symmetry mandates that their masses 
are generated by the VEV that breaks the $U(1)_{Z^\prime}$ 
gauge symmetry.  
In this paper we showed that the observed low scale 
gauge coupling parameters are also in good agreement with the
$Z^\prime$ model. The situation is in fact identical to that 
of the MSSM at one--loop level, whereas two--loop effects are small
and can be absorbed into the unknown mass thresholds. Kinetic mixing
effects are also small and can be neglected in the analysis. 
Above the intermediate breaking scale the weak hypercharge 
is embedded in a non--Abelian group and kinetic mixing cannot 
arise. Below the intermediate breaking scale kinetic 
mixing arises due to the extra pair of electroweak doublets, 
but it is found to be small and does not affect the results.
We further showed that the $Z^\prime$ model can indeed 
account for the observed signal, while providing for a rich
scalar sector that includes the Standard Model Higgs 
and the scalar resonance, as well as numerous other states 
that should be generated in the vicinity of this 
resonance. If such a resonance is observed 
in forthcoming data, future higher energy colliders
will be required to decipher the underlying physics. 
\section*{Note added:}

While this paper was under review the ATLAS and CMS collaborations
reported that
accumulation of further data did not substantiate the 
observation of the di--photon excess \cite{atlasaug,cmsaug}, 
indicating that the initial observation was a statistical fluctuation.
In our view, rather than being a negative outcome of the 
initial signal, it reflects the robustness and expediency
of collider based experiments, and we eagerly look forward for 
future such ventures. We further remark that while a di--photon 
excess at 750GeV was not substantiated by additional data, 
a di--photon excess at energy scales accessible at the 
LHC provides a general signature of the string derived 
$Z^\prime$ model of ref. \cite{frzprime}. We are indebted
to our colleagues in ATLAS and CMS, as well as those 
in ref. \cite{flurry} for drawing our attention to this 
possibility. Similarly, the gauge coupling 
analysis and the pertaining analysis 
that we presented in this paper is valid
for $Z^\prime$ and di--photon excess in the multi--TeV 
energy scale. 

\section*{Acknowledgments}

AEF thanks the theoretical physics groups at 
Oxford University and Ecole Normal Superier in Paris
for hospitality.
AEF is supported in part by the STFC (ST/L000431/1). 
LDR is supported by the "Angelo Della Riccia'' foundation.

\end{document}